\PassOptionsToPackage{pdfpagelabels=false}{hyperref}
\documentclass[useAMS,usenatbib]{mnras}
\pdfoutput=1

\usepackage{newtxtext,newtxmath}
\usepackage{newtxtext,newtxmath}

\usepackage[T1]{fontenc}
\usepackage{ae,aecompl}

\usepackage{amsmath}	
\usepackage{siunitx}
\usepackage[utf8]{inputenc}
\usepackage{xcolor}
\usepackage{times}
\usepackage{color}
\usepackage[pdftex]{graphicx}
\usepackage{natbib}
\usepackage{comment}
\usepackage[normalem]{ulem}
\usepackage{xspace}
\usepackage{rotating}
\usepackage{pdflscape}	

\def\hi{\ifmmode {\mbox H{\scshape i}}\else H{\scshape i}\fi\xspace}
\def\hii{\ifmmode {\mbox H{\scshape ii}}\else H{\scshape ii}\fi\xspace}
\def\h2{\ifmmode {\mbox H$_2$}\else H$_2$\fi\xspace}
\def\micron{\ifmmode {\mbox $\mu$m}\else $\mu$m\fi\xspace}

\newcommand{\dtg}{\ensuremath{{\rm DTG}}}
\newcommand{\dtm}{\ensuremath{{\rm DTM}}}

\title[Cosmic evolution of galaxy dust with metallicity]{Observed cosmic evolution of galaxy dust properties with metallicity and tensions with models
}

\author[]{Gerg\"o Popping$^{1}$\thanks{E-mail:gpopping@eso.org} \& C\'eline P\'eroux$^{1,2}$
\\
\\
$^{1}$European Southern Observatory, Karl-Schwarzschild-Str. 2, D-85748, Garching, Germany\\
$^{2}$Aix Marseille Universit\'e, CNRS, LAM (Laboratoire d'Astrophysique de Marseille) UMR 7326, 13388, Marseille, France \\
}
\date{Accepted XXX. Received YYY; in original form ZZZ}

\pubyear{}

\begin{document}
\maketitle

\begin{abstract}
The dust abundance of the interstellar medium plays an important role in galaxy physics, the chemical evolution of matter and the absorption and re-emission of stellar light. The last years have seen a surge in observational and theoretical studies constraining the dust-abundance of galaxies up to $z\sim5$. In this work we gather the latest observational measurements (with a focus on absorption studies covering metallicities in the range $6.8 < 12 + \log{(O/H)}<9$) and theoretical predictions (from six different galaxy formation models) for the dust-to-gas (\dtg) and dust-to-metal (\dtm) ratio of galaxies. The observed trend between \dtg\ and \dtm\ and gas-phase metallicity can be described by a linear relation and shows no evolution from $0<z<5$. Importantly, the fit to the \dtg-metallicity relation provides a refined tool for robust dust-based gas mass estimates inferred from millimeter dust-continuum observations.
The lack of evolution in the observed relations are indicative of a quickly reached balance (already when the Universe was 1.2 Gyr old) between the formation and destruction of dust and a constant timescale for star-formation at fixed metallicities over cosmic time. None of the models is able to reproduce the observed trends over the entire range in metallicity and redshift probed. The comparison between models and simulations furthermore rules out some of the current implementations for the growth and destruction of dust in galaxy formation models and places tight constraints on the predicted timescale for star-formation. 
\end{abstract}




\section{Introduction} \label{Introduction}

A fundamental component of galaxy formation is the cycle of baryons in
and out of galaxies. This picture emerges from both simulations and
observations alike \citep{peroux2020,Walter2020}. The flow of gas, metals,
and dust between stars, the interstellar medium, and circum-galactic
medium has a tremendous impact on galaxy evolution. The transfer of
metals between interstellar gas and the dust phase, in particular, constitutes an important component of this baryon cycle.

Interstellar dust is known to be made of small solid particles (grains distributed in size with an average of $\sim$1 $\mu m$) of carbon, silicates, and ices from other
chemical species \citep{draine2003}. Despite the small mass fraction in dust (1\%)
compared to gas in the interstellar medium of galaxies (99\%, e.g., \citealt{remy2014}), dust has a
tremendous impact in several fundamental physical processes.  Dust plays a key role in the radiative transfer, chemistry, and thermodynamics which
all profoundly impact galaxy evolution. In addition, a
substantial fraction of all the metals produced in the Universe are
locked into solid-phase dust grains (approximately 44 per cent for the Milky Way, based on \citealt{remy2014}).

Dust is critical in the thermal balance of gas as well as in shielding
the cores of dense clouds from ultra-violet radiation, allowing for the formation of
molecules which are critical to the star-formation process. The surfaces of dust
grains catalyse a range of chemical reactions that influence the
structure of interstellar medium and star formation \citep{Hollenbach1971, Hollenbach2012, Gong2017}. In
addition, they are responsible for the heating of the gas in
photodissociation regions by the photoelectric effect 
\citep{Draine1978}. These effects have poorly constrained  efficiencies and are highly environment dependent due to radiative transfer effects. Dust also acts as an efficient catalyst of
the formation of the most abundant molecule in the Universe, molecular hydrogen (H$_2$) in the interstellar medium \citep[e.g.,][]{Gould1963,Cazaux2009, Romano2022}. Indeed, giant molecular clouds composed of
molecular gas are the major formation sites of stars in galaxies
\citep[e.g.,][]{Blitz1993, Fukui2010}. In turn, dust absorption of
far-ultraviolet and optical photons can shape the temperature
structure of the neutral interstellar medium \citep{Goldsmith2001,Krumholz2011,Liang2019}.

Dust grains also play a crucial role in radiative processes within the
interstellar medium by absorbing ultraviolet and optical starlight and
reradiating it at far-infrared wavelengths. An important observational
consequence of this property is a induced change in the spectral
energy distributions of galaxies \citep{Calzetti2000, Buat2002, Salim2020}. As a consequence, our census of the
star-formation rate (SFR) density of the Universe with cosmic time is
highly incomplete unless we include the fraction of star formation
obscured by dust \citep[e.g.,][]{Steidel1999,Takeuchi2010,Kennicutt2012, Madau2014, Bouwens2020, Gruppioni2020, Fudamoto2021}.

Dust furthermore has an impact on gas dynamics in dusty clouds through radiation
pressure \citep[e.g.,][]{Ishiki2017}.  Moreover, the typical mass of
the final fragments in star-forming clouds is regulated by dust
cooling \citep{Whitworth1998, Omukai2000, Omukai2005, Larson2005, Schneider2006} with potential dramatic impact on the stellar initial mass function. For example, recent theoretical work suggests that the initial mass function in metal-poor (and dust-poor, $Z/Z_\odot \approx 10^{-5}$) environments is top heavy compared to a Chabrier \citep{chabrier2003} initial mass function \citep{chon2021}.

The ejection of dust from galaxies contributes to the metal enrichment
of the circumgalactic and ultimately intergalactic medium while providing an additional cooling
channel \citep{Ostriker1973, menard2010, peeples2014,peek2015,Vogelsberger2019, PerouxNelson2020, Wendt2021}. Eventually, dust coagulation in proto-planetary
discs becomes a key ingredient of planet formation \citep[e.g.,][]{Okuzumi2009, Kataoka2014}.

For all these reasons, a complete diagnostic of dust physical characteristics and their
relation to galaxy properties are essential. Establishing the galactic scaling relationships that apply to dust thus appears a natural tool. Specifically, the dust-to-gas ratio
(hereafter \dtg) quantifies the fraction of the interstellar mass
locked onto dust grains. It is defined as the dust mass divided by the
gas mass or alternatively dust surface density (or column density)
divided by the gas surface density (or column density). Likewise, the
dust-to-metals ratio (\dtm) represents the fraction of the metal mass
incorporated into the solid phase. The \dtg\ and \dtm\ ratios are fundamental parameters resulting from the interstellar gas-dust cycle, and are expected to substantially vary with environment and in particular gas metallicity \citep{lisenfeld1998, draine2007, galliano2008}. The relation between the \dtg\ and \dtm\ of galaxies as a function of their metallicity provides key constraints for the physics and timescales governing the buildup and destruction of dust in galaxies \citep[e.g.,][]{asano2013,remy2014,zhukovska2014, feldmann2015}.


The goal of this work is to provide a new appraisal of the dust
abundance in galaxies as a function of both galaxy properties
and cosmic time with the aim to assess their impact on our current
knowledge of galaxy evolution. To this end, we have gathered the
latest observational measurements and results from a selected sample
of state-of-art semi-analytical and hydrodynamical simulations. A broad comparison between observations and simulations is very timely, as in the last five years significant advancements have been made in  constraining the \dtg\ and \dtm\ of galaxies at $z>0$ through absorption and emission studies \citep{decia2016,wiseman2017,peroux2020, shapley2020} and modeling these properties in cosmological simulations \citep[e.g.,][]{Mancini2015,popping2017,Ginolfi2018, mckinnon2018,li2019,triani2020,vijayan2019,hou2019,graziani2019, Kannan2021, Dayal2022}.

This paper
is organised as follows. In Section \ref{Observation} and Section \ref{Sec:models} we present the observations and simulations used in this work, respectively. The comparison between the two is presented in Section \ref{sec:results}, whereas we discuss and summarise the results in Sections \ref{sec:discussion} and \ref{sec:summary}. We adopt the following cosmological parameters (Planck Collaboration et al. 2016): H$_0$ = 67.74 km s$^{-1}$ Mpc$^{-1}$, $\Omega_M$ =
0.3089, and $\Omega_{\Lambda}$ = 0.6911.

\section{Observational compilation} 
In this Section we present the data compilation used in this work to explore the relation between \dtg\ and \dtm\ of galaxies with metallicity and their cosmic evolution. We first discuss the compilation of local galaxies with \dtg\ and \dtm\ information based on their emission (Section \ref{low-z_obs}) and then discuss the $z>0$ compilation based on absorption measurements (Section \ref{high-z_obs}).

\label{Observation}

	\begin{figure*}
	\centering
	\includegraphics[width=1.\textwidth]{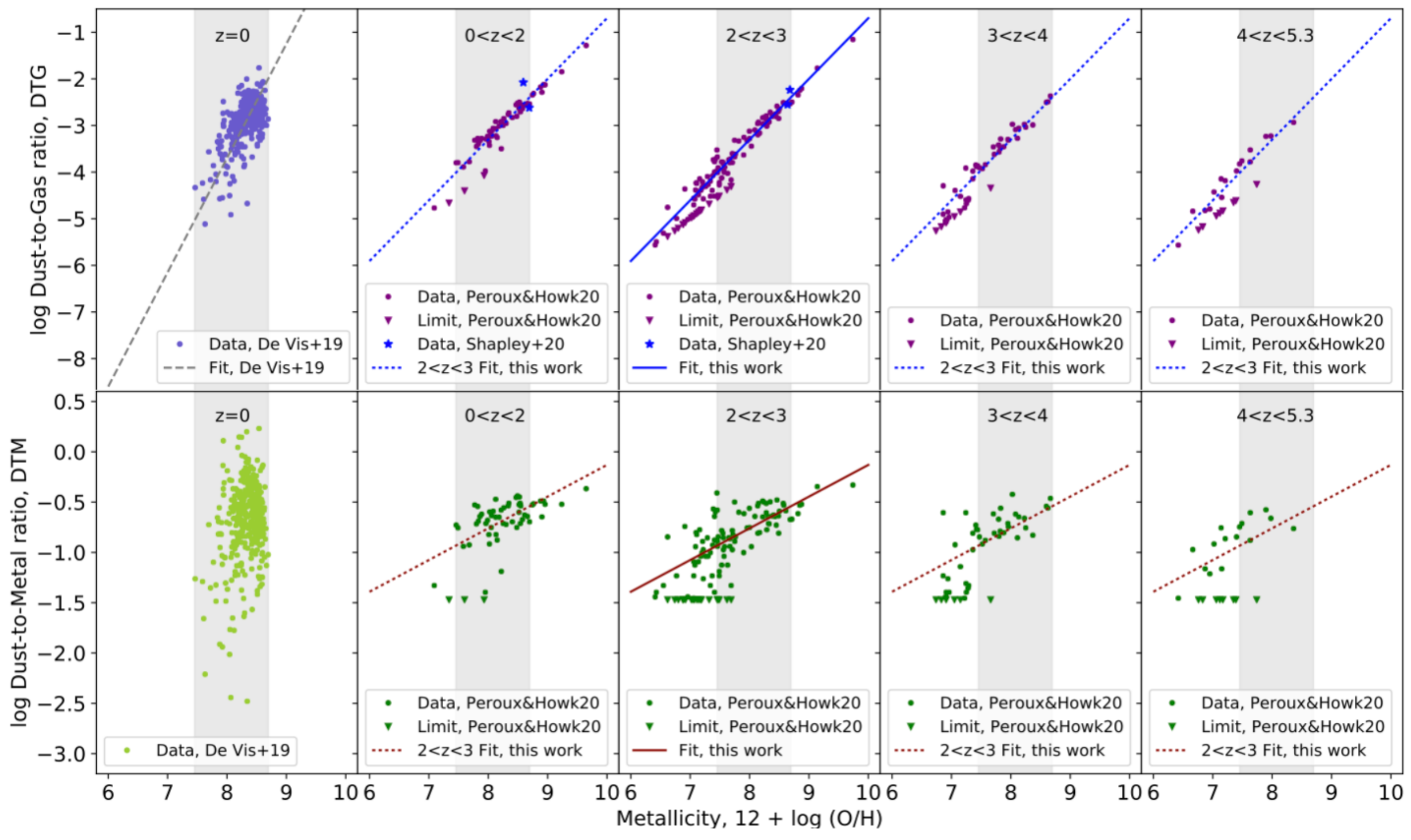}
	\caption{Observed cosmic evolution of the dust-to-gas and dust-to-metal ratios with metallicity. The grey band reproduced in all panels marks the span in metallicity of the z$=$0 data sample (detected through their emission). The z$>$0 observations (based on absorption studies) extend to both lower and higher metallicities. The DTG are tightly related to metallicity over almost 4 dex as indicated by the small scatter around the fitted linear regression. Furthermore, the fit to the observational data indicate little evolution with cosmic time of that relation at z$>$0. The scatter in the DTM measurements is larger, still providing little support for a redshift evolution of the DTM-metallicity relation. 
	\label{fig:obs}}
	\end{figure*}

\begin{table}
\caption{Linear regression fits to the observed DTG and DTM evolution with metallicity plotted on Fig~\ref{fig:obs}. Note that the DTM and DTG are correlated such that DTM $=$ DTG / Z, where Z is the gas-phase metallicity expressed as $M_{\rm metal} / M_{\rm gas}$.}
\centering
\begin{tabular}{ccccc}
\hline
& Redshift	&	Slope &Intercept &Reference 	\\	
\hline	
$\log$ DTG & z=0    &2.45   &-23.30 &\citet{de-vis2019}\\
    &z$>$0  &1.30   &-13.72 &this work\\
$\log$ DTM &z$>$0  &0.32   &-3.28 &this work\\
\hline
\end{tabular}
\label{tab:obs_fit}
\end{table}

\subsection{Local Measurements of DTG \& DTM}
\label{low-z_obs}

The \dtg\ and \dtm\ values are well
constrained in our Galaxy, based on modeling the infrared emission
and optical/ultraviolet extinction \citep[e.g., ][]{draine2007}. Modern surveys provide fresh measurements of these quantities in
galaxies beyond the Milky Way \citep{lianou2019, vilchez2019}. A remarkable
coordinated effort undertaken under the DustPedia collaboration has
assembled observational constraints on DTG and DTM from a large sample of 
nearby galaxies \citep[e.g.,][]{remy2014, de-vis2019, galliano2021}. In this work, the dust masses are estimated from a Spectral Energy Distribution (SED) fit to the galaxy photometry. The total gas mass is the sum of direct HI-emission mass measurements and H$_2$ masses estimated through CO observations \citep{remy2014} or estimated assuming H$_2$-to-HI and HI-to-stellar mass ratios \citep{de-vis2019}. The gas-phase oxygen abundances are derived using multiple strong emission-line calibrations (using the \citealt{Pilyugin2016} S-calibration, based on $N_2$, $R_2$ and $S_2$). We note that metallicities based on strong emission-line calibrations have been reported to present significant dispersion depending on the calibration used \citep[up to 0.3--0.4 dex][]{kewley2008, scudder2021}. All these quantities are integrated over the whole galaxy. 
The resulting data are reproduced in the left panels of
Fig.~\ref{fig:obs}. The data indicate
that galaxies of lower metallicities have a lower \dtg\ and \dtm\ than
more metal-rich galaxies.

\subsection{Measurements of DTG \& DTM at z$>$0}
\label{high-z_obs}

Measurements of the \dtg\ and \dtm\ of galaxies from their emission have so far mostly been limited to the local Universe (but see \citealt{shapley2020}). 
A new approach based on dust depletion has enabled the efficient measurement of the \dtg\ and \dtm\ of galaxies beyond the local Universe. This empirical method has been developed to derive the dust depletion level in the Milky Way based on gas-phase element abundances in the local part of our Galaxy \citep{jenkins2009}. The information from this study reveals the relative proportions of different elements to be incorporated into dust
at different stages of grain growth. This method uses observed abundance patterns with assumptions about relative abundances to define depletion sequences, where the depletion effects on gas-phase abundances of all the elements is described by a single parameter. This work has been subsequently extended to both the Magellanic clouds using dedicated HST observations of the SMC \citep{roman-duval2014, jenkins2017} and the LMC \citep{roman-duval2019, roman-duval2021, roman-duval2022}.

This simple scheme can be similarly implemented to derive the dust contents and metallicities of absorption-line systems that are seen in the spectra of distant quasars. Observing the gas in absorption against bright background sources provides a powerful technique to assess the gas, metal and eventually dust content of galaxies \citep{peroux2007, quiret2016,poudel2017, decia2018a, berg2021}. In these quasar absorbers, the minimum gas density detectable is set by the brightness of the background source. The detection efficiency is thus independent of cosmic time \citep{peroux2002}. The high-column density gas probed in absorption is known to be strongly associated with galaxies. Multiple observational results based on Integral Field Spectroscopy have shown that the quasar absorbers are also tracing overdensities of galaxies down to low-luminosities (L = 0.1 L$_*$) \citep[see e.g.][]{peroux2019, hamanowicz2020, Dutta21b}. While the physical properties of absorbing-galaxies remain to be measured statistically, early results indicate that their stellar masses range from M$_*$=10$^8$-10$^{11}$ M$_\odot$ \citep{augustin2018}.

In large samples, the pencil-approach provides constraints on the physical properties of the average galaxy population. Absorption lines provide a measure of the surface density or column density of atoms, ions or molecules between the observer and the background source (expressed in atoms cm$^{-2}$). Absorption techniques directly count the number of atoms in a given phase of the gas. When probing regions dominated by neutral atomic gas, the metallicity measurements trace only the dominant ionisation states of the metal element and hydrogen. Additionally, absorption-based gas metallicity measurements are weakly dependent on the excitation conditions making them robust estimators in comparison with emission-based measurements \citep{kewley2019, maiolino2019}. 

\cite{jenkins2009} provides a straightforward method to infer the level of loss of metals into dust grains, allowing one to determine the total elemental abundances to be derived from the measured gas-phase column
densities. Specifically, \cite{decia2018a}  determined corrections for the zero depletion element abundances caused by nucleosynthesis
effects in such quasar absorbers, with some additional guidance from
the trends seen in metal-poor stars in our Galaxy. 

By using the dust sequences to assess the depletion $\delta$ for each of the elements, one can ultimately calculate their sum, the total \dtg\ ratio for each quasar absorber \citep{decia2013, decia2016, wiseman2017}. These results, augmented with the latest few measurements, are available in an updated compilation in Supplementary Table 4\footnote{https://www.annualreviews.org/doi/suppl/10.1146/annurev-astro-021820-120014} of \cite{peroux2020}. These observations are also plotted in the four panels to the right of the first panel in the top row  of Fig~\ref{fig:obs} for the following redshift ranges: 0$<$z$<$2, 2$<$z$<$3, 3$<$z$<$4 and 4$<$z$<$5. Most of the observations lie in the 2$<$z$<$3 interval, where abundance determination in quasar absorbers can be robustly made at optical wavelengths observed from the ground. Upper limits represent absorbers in which carbon is the only significant contributor to the dust content. Equally, the dust-to-metal ratio for an individual element, $X$, is related to its depletion: $\dtm = 1 - 10^{\delta}$ \citep[e.g.,][]{vladilo2004,decia2016,peroux2020}. These observations are displayed in four panels to the right of the bottom panel of Fig~\ref{fig:obs}.

\subsection{Evolution of DTG \& DTM extended to low-metallicity}
Observations indicate that both \dtg\ and \dtm\ ratios are a
strong function of metallicity. This result was determined based on various studies focusing on unresolved  \citep{Issa1990, lisenfeld1998, Hirashita2002, draine2007, Galametz2011, remy2014,de-vis2019} and resolved  \citep{giannetti2017,chiang2018} nearby galaxies, as well as absorption based studies focusing on high-redshift objects \citep{decia2013, decia2016, decia2018, zafar2013a, sparre2014, wiseman2017}. Importantly, the quasar absorbers depletion measurements
probe a new regime of low-metallicity gas which remained inaccessible
in nearby galaxies (the probed metallicity range at $z=0$ is narrower that at $z>0$, as indicated by the grey band in Figure~\ref{fig:obs}). 

Our results support the conclusion that the \dtm\ and \dtg\ ratios are a strong function of metallicity. We find that at $z>0$ the \dtg\ of the absorbers increases with approximately 4 dex in the metallicity range $7 < 12 + \log{(\rm{O/H})} < 9$, while the \dtm\ increases with approximately 1 dex over the same metallicity range.

Globally, a larger scatter in \dtg\ is observed at 12+log(O/H)$<$8, indicating a
change in the dust assembly processes at low metallicities. These
findings show compelling evidence at both low
and high redshifts that the \dtg\ and \dtm\ in galaxies are not
constant, but instead are a strong function of galaxy physical
property as traced for example by metallicity.

\subsection{Cosmic Evolution of DTG \& DTM}

High-redshift measurements provide a unique
opportunity to evaluate the evolution of the \dtg\ and \dtm\ relations
with metallicity as a function of cosmic times. As detailed in Sections~\ref{low-z_obs} and \ref{high-z_obs},  observed quantities are derived from different techniques at z=0 and z$>$0. In particular, at $z=0$ the metallicities correspond to the integrated ionised gas based on strong emission-line calibrations, whereas at $z>0$ the metallicities are based on pencil beam absorption abundance determinations of the neutral gas. Some systematic differences between the two approaches are thus expected, although several analyses report that abundances from emission and absorption only vary by a maximum of $\sim{}0.6$ dex \citep{rahmani2016}. 

In spite of these differences, it is noteworthy that the neutral gas measurement follows the trends seen in low-redshift galaxies at high metallicities. There is little evidence that the \dtg\ and \dtm\ behave differently with redshift. In the absence of a physical motivation for a specific functional form, we fit the observations with a simple
power-law function as represented by the lines in Figure \ref{fig:obs}.  It is remarkable to see that the \dtg\ and \dtm\ as measured through dust depletion from $z=0$ and $z=5$ are so well described by the simple power-law function (the results of the fits are tabulated in Table \ref{tab:obs_fit}). This lack of evolution can be naturally explained as galaxies evolve along the \dtg\ and \dtm\ vs. metallicity relationship as a function of cosmic time.

\begin{landscape}
\begin{table}
  \centering
  	\begin{tabular}{c c p{20mm} p{30mm} c p{25mm} p{25mm} p{25mm}} 
		\hline
		Model & Simulation type & Condensation & Dust yields & Grain-growth model & Time scale for grain-growth depends on & Gas mass cleared of dust per SN as defined in \citet{mckee1989} & Star formation recipe \\
		\hline
		\citet{popping2017} & SAM & AGB, SN Type Ia and II  & fixed fraction of metals: 0.15 (SN) and 0.2 (AGB) & \citet{zhukovska2014} &  metallicity, \newline  gas surface-density &  fixed mass per SN (600 (carbonaceous) and 980 (sillicates)) $\rm{M}_\odot$& star-formation efficiency evolves with H$_2$ surface density \citep{popping2014}\\
		&&&&&&&\\
		&&&&&&&\\
		\citet{vijayan2019} & SAM & AGB, SN Type II & fixed fraction for SN (0.15) and AGB yields from \citet{ferrarotti2006} & \citet{zhukovska2014} & dust abundance & fixed mass per SN (1200 $\rm{M}_\odot$)& star-formation efficiency evolves with gas mass and galaxy disc dynamical time \citep{henriques2015}\\
		&&&&&&&\\
		&&&&&&&\\
		\citet{triani2020} & SAM & AGB, SN Type II & fixed fraction of metals: 0.15 (SN) and 0.2 (AGB) & \citet{dwek1998} & metallicity & scales negatively with metallicity \citep{yamasawa2011} & scales linearly as a function of H$_2$ surface density \citep{triani2020}\\
		&&&&&&&\\
		&&&&&&&\\
		\citet{hou2019} & hydrodynamical& SN Type II  & yield tables from \citet{nozawa2006} & \citet{hirashita2011} & metallicity and \dtg\ (with fixed dense gas fraction) & ISM density (see \citealt{aoyama2017})&scales linearly with local gas density \citep{aoyama2017}\\
		&&&&&&&\\
		&&&&&&&\\
		\citet{li2019} & hydrodynamical&AGB, SN Type II & fixed fraction of metals: 0.15 (SN) and 0.2 (AGB)  & \citet{dwek1998} & temperature, density,\newline metallicity & fixed mass per SN ($\sim$2040 $\rm{M}_\odot$) &scales with H$_2$ density and disc dynamical time \citep{dave2019}\\
		&&&&&&&\\
		&&&&&&&\\
		\citet{graziani2019} &hydrodynamical& AGB, SN Type II & yield tables from \citet{Schneider2004} and \citet{Bianchi2007} (SN) and \citet{ferrarotti2006} and \citet{Zhukovska2008} (AGB) & \citet{Bennassuti2014} & temperature, density,\newline metallicity & fixed mass per SN ($\sim$2040 $\rm{M}_\odot$)&scales with local gas density \citep{springel2005b}\\
		\hline
	\end{tabular}
	\caption{Overview of the implementation of dust-chemistry of the various models discussed in this work. All these properties contribute to the physics encoded in the relation between \dtg\ or \dtm\ and gas-phase metallicity takes. 
	\label{tab:model_overview}}
\end{table}
\end{landscape}

\section{Model descriptions}
\label{Sec:models}
We now compare the \dtg\ and \dtm\ derived from observations to a number of cosmological semi-analytic and hydrodynamical simulations of galaxy formation and evolution. These simulations were selected to include the tracking of dust production and destruction for a full cosmological volume over a large redshift range (partially) overlapping with the observational sample. The predictions by the simulations by \cite{mckinnon2018} were unfortunately not available to us. Predictions by the \citet{Kannan2021} model focus on redshift $z>5.5$, for which there are no observations to compare to yet.

All discussed models include typical processes that control the dust enrichment of the ISM, such as the condensation of metals in the stellar ejecta of SNe and AGB stars, dust mass growth through the accretion of metals onto dust grains, the destruction of dust in the vicinity of SNe due to shocks (accounting for the effects of correlated SNe: SNe exploding in existing super-bubbles created by previous SNe in the association) and the destruction of dust in hot gas due to thermal sputtering. Since the reversed shock in supernova acts on scales smaller than resolved in these simulations, the impact of the reversed shock on dust yields is taken into account by an effective yield for SNe (rather than the yield before dust is destroyed by a reversed schock). Furthermore they all include astration (dust that is destroyed while it is participating in the formation of new stars) and the flows of dust into and out of galaxies due to stellar and AGN winds and cold gas accretion. None of these models explicitly model the surviving rate of dust in stellar and AGN winds. 

Below we provide a brief overview of the details of dust modelling in  the various models considered in this work and we provide a summary of the relevant parameters in Table~\ref{tab:model_overview}. We refer the reader to the original papers for a detailed description.

\subsection{Semi-analytic models}
\subsubsection{\citet{popping2017}}
The dust-evolution model of \citet{popping2017} builds upon the \texttt{SANTA CRUZ} semi-analytical model \citep{somerville1999,somerville2008, somerville2015,popping2014}. The model includes the condensation of dust in the ejecta of Type Ia and II supernovae and asymptotic giant branch (AGB) stars. The model adopts the prescription from \citet{zhukovska2014} to model grain growth in the molecular ISM with a timescale that depends on the gas phase metallicity and surface density of the dense ISM. Grain-growth only takes place in the molecular phase of the ISM, calculated using the \citet{gnedin2010} partitioning recipe. The destruction of dust is modelled following \citet{dwek1980} and \citet{mckee1989} assuming a fixed mass of gas cleared of dust per SN event. The model furthermore includes astration, thermal sputtering of dust and the in- and outflow of dust into the ISM.

\subsubsection{\citet{vijayan2019}}
The dust chemistry model presented in \citet{vijayan2019} is implemented within the \texttt{L-GALAXIES} semi-analytic model \citep{henriques2015}. The model includes the condensation of dust in SN II and AGB stars, grain-growth in the ISM, destruction of dust by SNe, astration and the in- and outflow of dust into the ISM. \citet{vijayan2019} assume a timescale for the accretion of metals onto dust-grains in the molecular ISM that scales with the dust-abundance of the cold gas, calibrated to match observed dust masses at high redshifts. The molecular phase of the ISM is calculated using the \citet{mckee2009}  partitioning recipe for atomic and molecular gas. The destruction of dust follows the work of \citet{dwek1980} and \citet{mckee1989}, assuming a fixed mass of gas cleared of dust per SN event. In the \citet{vijayan2019} model all dust that is transferred to the hot halo is destroyed directly.

\subsubsection{\citet{triani2020}}
\citet{triani2020} implemented a detailed dust prescription in the Semi-Analytic Galaxy Evolution (\texttt{SAGE}, \citealt{croton2016}) model. This version of the model, called \texttt{DUSTY SAGE}, includes condensation of dust in the ejecta of Type II supernovae and AGB stars, grain growth in molecular clouds, destruction by supernovae shocks, and  the  removal  of  dust  from  the  ISM through astration,  reheating, inflows, and outflows \citep{triani2020}. The authors adopt the procedure presented in \citet{dwek1998} to model grain growth in the molecular ISM. The molecular fraction of the ISM is calculated using the \citet{blitz2006} partitioning recipe for atomic and molecular gas. The destruction of dust by SNe is modeled following \citet{dwek1980} and \citet{mckee1989}, assuming that the mass of dust swept up per SN event depends on the gas-phase metallicity (following \citealt{yamasawa2011}), such that more dust is swept up in metal-rich environments.

	\begin{figure*}
	\centering
	\includegraphics[width=1.\linewidth]{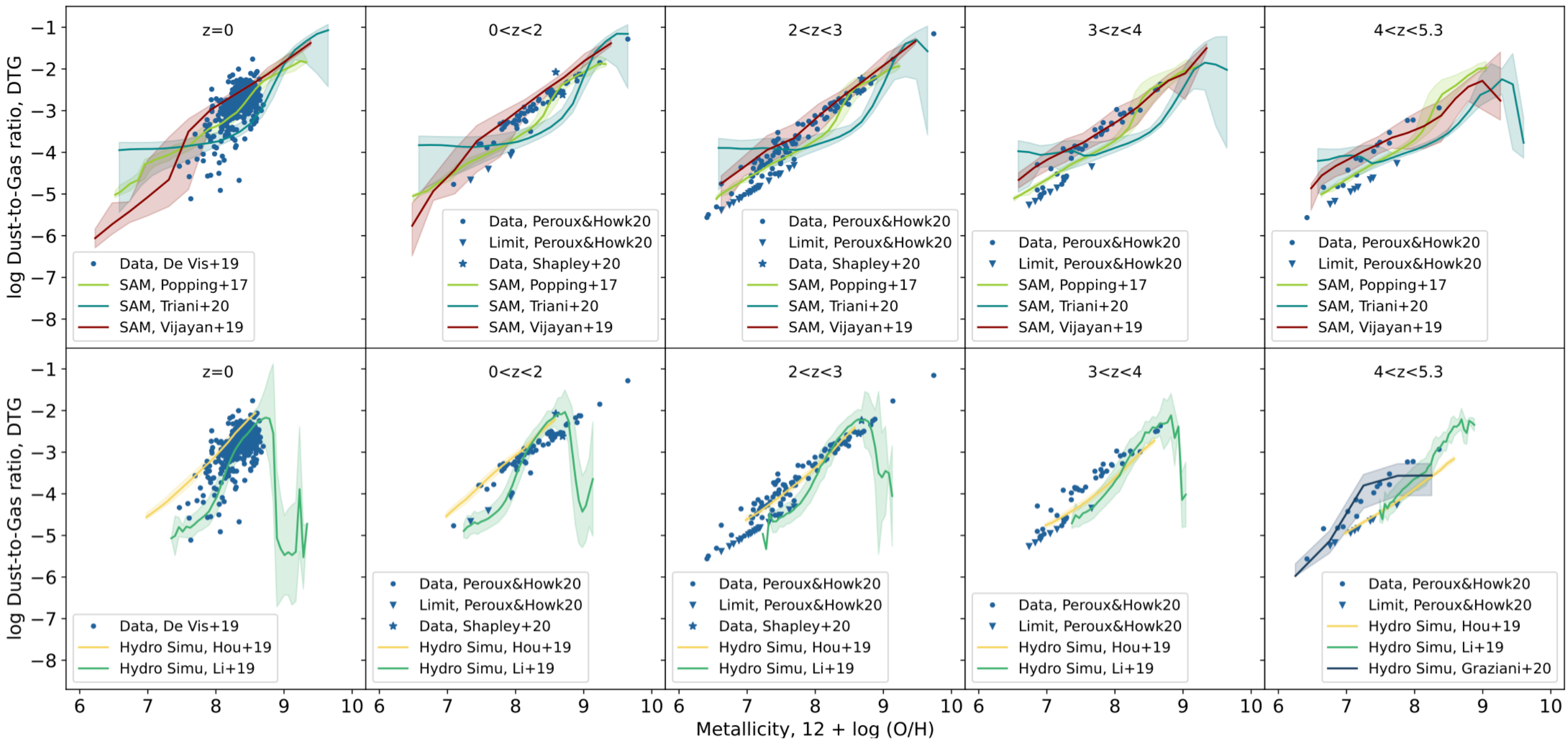}
	\caption{Observed and modeled evolution of dust-to-gas ratios with metallicity. The data points are identical to the top row of Figure~\ref{fig:obs}. The lines in the top row of this figure depict three different semi-analytical models with associated 1-$\sigma$ error estimates, while the lines in the bottom row depict three different hydro-dynamical models. The models reproduce the observed DTGs with varying degree of success.	\label{fig:dtg} }
	\end{figure*}

\subsection{Hydrodynamical models}
\subsubsection{\citet{hou2019}}
\citet{hou2019} presents predictions for the evolution of \dtg\ and \dtm\ by making use of the cosmological simulation presented in \citet{aoyama2017} and \citet{hou2017}, a modified version of the \texttt{GADGET-3} $N$-body/smoothed particle hydrodynamic code presented in \citet{springel2005}. The model considers dust production in Type-II SNe and the accretion of metals onto dust gains in the dense ISM following \citet[similar to \citealt{dwek1998}]{hirashita2011}. The timescale for dust accretion depends on the metallicity and \dtm\ of the gas. Dust accretion only occurs in the dense ISM and it is assumed that the ISM has a fixed fraction of dense gas of 0.1. The destruction of dust by SN shocks is modeled following \citet{mckee1989}, where the gas mass swepped up per SN event depends of the gas density of the particle. Additionally, dust is destroyed by thermal sputtering. 

The \citet{hou2019} model is the only one presented in this work that includes a dust size distribution and divides dust in two type of grains, large and small. Small and large grains have different efficiencies for the accretion of metals onto dust grains and the destruction of dust by SN shocks. Furthermore, the model includes the coagulation of small grains and the shattering of large grains (see also \citealt{li2020}). 

\subsubsection{\citet{li2019}}
\citet{dave2019} includes the tracking of dust production and destruction as a part of the \texttt{SIMBA} cosmological simulation, built upon the \texttt{GIZMO} cosmological gravity plus hydrodynamic solver \citep{hopkins2015}. \citet{li2019} presents the evolution of the \dtg\ and \dtm\ predicted by the \texttt{SIMBA} simulation. The model for dust production and destruction includes the condensation of dust in the ejecta of AGB stars and Type-II SNe and the accretion of metals onto dust grains \citep[following][]{dwek1998} assuming a accretion timescale that depends on the temperature, metallicity and density of the ISM. Metal accretion can in theory take place in every particle in the simulation, although it will be less efficient in the diffuse ISM due to the density dependence. The model furthermore includes thermal sputtering of dust and the destruction of dust in SN blast waves \citep{dwek1980,mckee1989}, assuming a fixed mass of gas swept up per SN event. Dust is additionally destroyed in hot winds, due to X-ray heating and through astration.

\subsubsection{\citet{graziani2019}}
\citet{graziani2019} presents \texttt{DUSTYGADGET}, a code following the evolution of dust grains in different phases of the ISM and the spreading of dust and atomic metals by galactic winds throughout the circum- and intergalactic medium. This code is built upon the \texttt{GADGET} \citep{springel2005} model and extensions presented in \citet{Tornatore2007a,Tornatore2007b} and \citet{Maio2009}. The model follows the condensation of dust in population III SN, population II/I core-collapse SNe and AGB stars using yield tables from \citet{Schneider2004}, \citet{Bianchi2007} and \citet{ferrarotti2006} and \citet{Zhukovska2008}, respectively. This is the only of the discussed models that adopts dust-yield tables for both SNe and AGB stars, describing the condensation of dust as a function of the stellar mass and metallicity, rather than a fixed condensation efficiency independent of stellar properties. The \citet{graziani2019} model describes the accretion of metals onto dust grains assuming a timescale that depends on the density, metallicity and temperature of a star-forming SPH particle and the destruction of dust by supernova shocks following \citet{mckee1989} with a fixed mass of dust cleared from the ISM per SN (with varying efficiencies for core-collapse and pair-instability SNe). Metal accretion can only take place in cold and star-forming particles, which is different from the implementation of \citet{li2019}. Lastly, the model includes grain sputtering in the hot plasma and astration. The \citet{graziani2019} model is evolved only down to $z=4$ due to computational limitations.

\begin{figure*}
	\centering
	 \includegraphics[width=1.\linewidth]{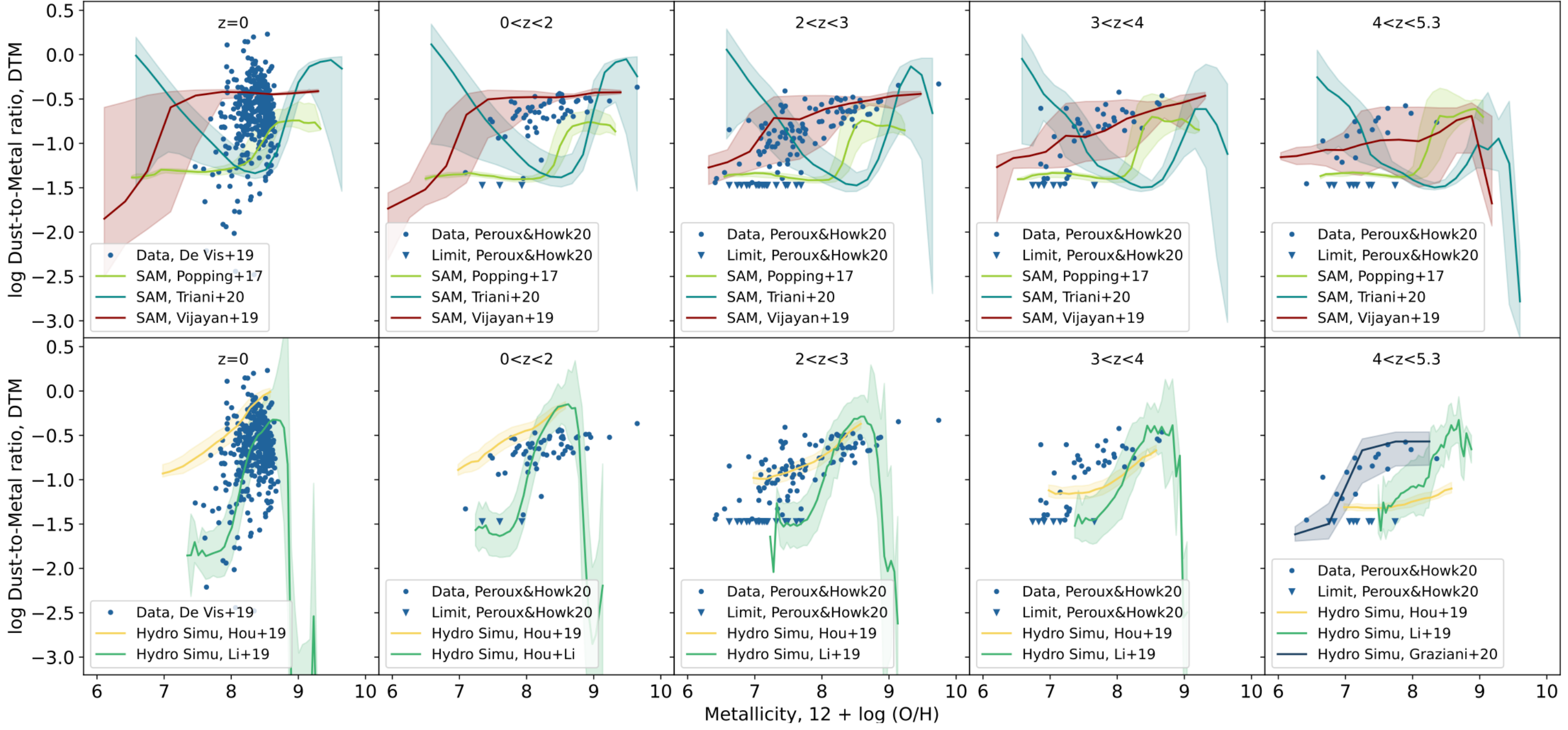}
	 \caption{Observed and modeled evolution of dust-to-metal ratios with metallicity. The data points are identical to the bottom row of Figure~\ref{fig:obs}. Similar to Figure \ref{fig:dtg}, the lines in the top row depict three different semi-analytical models with associated 1-$\sigma$ error estimates, while the lines in the bottom row depict three different hydro-dynamical models. The DTMs are reproduced by the models with a varying degree of success with vastly different predictions at fixed redshifts and for the cosmic evolution of the relation between \dtm\ and metallicity.	\label{fig:dtm} }
\end{figure*}

\section{Confronting data to models}
\label{sec:results}
\subsection{Dust-to-gas ratio }
Figure 	\ref{fig:dtg} shows the observed \dtg\ of galaxies as a function of their gas-phase metallicity (as presented in the top row of Figure~\ref{fig:obs}). The different columns represent different redshift bins, whereas the various lines mark the predictions by the respective models discussed in Section \ref{Sec:models}. The top row of Figure \ref{fig:dtg} shows a comparison between observations and model predictions by semi-analytic models, whereas the bottom row shows the comparison with hydrodynamic models.

It is remarkable that despite the different implementations between the various models for baryonic physics (e.g., the formation of stars, the accretion of gas onto galaxies, the growth of black holes and stellar and AGN feedback), the predicted relation between \dtg\ and gas-phase metallicity is roughly consistent between many of the models.  The \citet{popping2017}, \citet{triani2020}, \citet{li2019} and \citet{graziani2019} models predict a relation between \dtg\ and gas-phase metallicity that takes an 's-shape'. The \citet{hou2019} and \citet{vijayan2019} models on the other hand predict a roughly linear relation. The quick rise in \dtg\ around $12 + \log{(O/H)} \sim 8$ (8.5 for \citealt{triani2020}) is driven by the onset of accretion of metals onto dust grains as an important channel for the growth of dust mass (see for example \citealt{popping2017} and \citealt{triani2020}).  Furthermore, except for \citet{hou2019}, all other models discussed in this work predict that the relation between \dtg\ and metallicity is roughly constant with time. A constant relation between gas-phase metallicity and \dtg\ is in apparent agreement with observations \citep[Figure \ref{fig:obs},][]{decia2016,peroux2020, shapley2020}. The \citet{hou2019} model predicts that the normalization of this relation increases with cosmic time. 

The models discussed in this work reproduce the observed relation between \dtg\ and gas-phase metallicity with a varying degree of success. Although all models reproduce part of the probed dynamic range in metallicicy, a close look at the individual models does show differences. The \citet{triani2020} model predicts a \dtg\ at metallicities around $12 + \log{(O/H)} \sim 8-9$ that is about half a dex lower than the observations, independent of redshift. At lower metallicities the \citet{triani2020} model predicts a \dtg\ that is up to a dex too high. The \citet{popping2017} and \citet{li2019} models predict a \dtg\ at $12 + \log{(O/H)} \sim 8$ that is $\sim 0.25$ dex lower than the observations. The \citet{vijayan2019} model reproduces the \dtg\ of galaxies with a gas-phase metallicity larger than $12 + \log{(O/H)} =8$ well, but predicts a \dtg\ that is about half a dex too large at lower metallicites. The predictions by the \citet{hou2019} model for the relation between \dtg\ and gas-phase metallicity agree well with the observations for galaxies at $z<3$, but the predicted \dtg\ is up to a 0.5 - 1 dex too large at higher redshifts. The predictions by \citet{graziani2019} model at $z>4$ are in excellent agreement with the observed \dtg, but  the non detections fall outside of the typical one-sigma scatter predicted by this model. The \citet{li2019} and \citet{triani2020} models both show a steep drop in the \dtg\ for galaxies at the highest metallicities, corresponding to quenched galaxies not included in the observational sample.


\subsection{Dust-to-metal ratio }
In Figure~\ref{fig:dtm} we present the relation between the \dtm\ and metallicity of galaxies at various redshifts. We find significant differences in the predictions by the various models for the relation between \dtm\ and gas-phase metallicity. The \citet{popping2017}, \citet{vijayan2019} and \citet{li2019} models all predicts a clear 's-shape'. Similar to the \dtg\, the quick rise in \dtm\ at a metallicity of $12 + \log{(O/H)} \sim 8$ is driven by the onset of accretion of metals onto dust grains as a channel for dust formation. The \citet{graziani2019} model also predicts an 's-shape' at $z>4$, but the quick rise in \dtm\ already occurs at a metallicity of $12 + \log{(O/H)} \sim 7$. The \citet{hou2019} model predicts a roughly linear relation between \dtm\ and gas-phase metallicity at $z<2$ and an exponential relation at higher redshifts. The \citet{triani2020} model on the other hand predicts that the \dtm\ of galaxies decreases with increasing gas-phase metallicity up to $12 + \log{(O/H)} =8$ and increases again at higher metallicities. 

Besides differences in the shape of the relation between gas-phase metallicity and \dtm\, the various models also predict a different redshift evolution. The relation predicted by \citet{popping2017}, \citet{triani2020} and \citet{li2019} is fairly constant with time, whereas the \citet{vijayan2019} and \citet{hou2019} models predict that the relation between gas-phase metallicity and \dtm\ becomes steeper as a function of cosmic time.

We now continue to review the successes and shortcomings of the various models when comparing their predictions to the observations. The \citet{popping2017} model predicts a \dtm\ that is up to $\sim0.75$ dex too low, independent of gas-phase metallicity. Similar to the observations, the \citet{popping2017} predicts a quick rise in \dtm\, but at a metallicity of $12 + \log{(O/H)} \sim 8$, rather than a metallicity of $12 + \log{(O/H)} \sim 7$ as suggested by the observations. The \citet{triani2020} model predicts a decrease in \dtm\ as a function of gas-phase metallicity at $12 + \log{(O/H)} <8$, whereas the data suggests an increase in \dtm. At higher metallicities the \citet{triani2020} model predicts a \dtm\ that is between one and 0.2 dex lower than the observed \dtm. The predictions by the \citet{vijayan2019} model are in good agreement with the observations at $z>2$. At redshifts $z<2$ this model predicts a \dtm\ that is about 0.25 dex too high for galaxies with $12 + \log{(O/H)} >8$, suggesting the quick rise in \dtm\ occurs at too low metallicities. The \citet{li2019} model reproduces the observations fairly well at $z<2$, but predicts a \dtm\ that is typically too low at higher redshifts (by  0.5-0.75 dex). The predictions by \citet{hou2019} model are in decent agreement with the observations at $2<z<3$. At redshifts $3<z<5$ the predicted \dtm\ is up to 0.5 dex too low, whereas at $z<2$ the predicted \dtm\ is up to 0.5 dex too large. The predictions by the \citet{graziani2019} model at $z>4$ are in good agreement with the observed \dtm.

\subsection{Differences between models explained}
\label{sec:differences}
There are clear differences between the various model predictions and large discrepencies between the model predictions and observations. None of the presented models succeeds in simultaneously reproducing the relation between \dtg\ or \dtm\ and gas-phase metallicity over the full redshift and metallicity range covered by the observations. Overall, this suggests that the \dtg\ and \dtm\ of galaxies are an excellent test for the success of a model that follows the formation and destruction of dust. The differences between models also provide relevant clues to how various implementations of dust physics control the buildup of dust in galaxies.

The various models discussed in this work all predict a different critical metallicity at which the \dtg\ and \dtm\ rapidly increase. For some models, this critical metallicity evolves with redshift. The critical metallicity is set by the efficiency of dust growth \emph{and} inversely by the star-formation timescale of galaxies (\citealt{asano2013}, \citealt{zhukovska2014}, see also \citealt{feldmann2015}). When the timescale for star formation is longer, metals have more time to accrete onto dust grains until the metallicity increases to a higher value due to star-formation driven metal enrichment. The recipes employed for dust growth in the ISM between the different models discussed in this work are all similar, which suggests that the change in critical metallicity between models (and even with redshift for the individual models) are driven by changes in the timescale of star-formation. It is beyond the scope of this work to reconstruct the timescale for star-formation in every model as a function of galaxy properties and time, however, we do note that the various models all adopt different recipes for star-formation that result into different timescales (up to an order of magnitude, see the last column of Table~\ref{tab:model_overview}). Interestingly, the \citet{popping2017}, \citet{triani2020} and \citet{li2019} models all predict a critical metallicity at which the \dtg\ and \dtm\ rapidly increases that is higher than suggested by the observations (indicative of a star-formation time scale that is too short) and these are the only three models discussed in this work that adopt a star-formation recipe directly tied to the molecular gas budget in galaxies.

Besides the timescale of star-formation, there are a number of additional physical drivers for the differences in the predicted \dtm\ and \dtg. \citet{hou2019} predict a redshift evolution in the relation between gas-phase metallicity and \dtg\ and \dtm. This evolution is likely driven by their assumption of a dense gas fraction of 10\%, whereas the other models take the molecular hydrogen fraction of the cold gas as the dense fraction (which is thought to be of the order 50\% and higher for massive galaxies at early times, e.g., \citealt{lagos2011,fu2012,popping2014, popping2015}). By adopting a dense gas fraction of 0.1, only small amounts of dust can be formed through the accretion of metals onto dust grains. This naturally slows down the buildup of dust in the ISM at early times, resulting in a strong evolution in the \dtg\ at lower redshifts, rather than reaching an equilibrium between grain-growth and destruction early on (as in for example \citealt{popping2017}).

The \citet{vijayan2019} model also predicts a strong redshift evolution in the relation between \dtm\ and gas-phase metallicity (this evolution is not so obvious in the relation with \dtg). \citet{vijayan2019} adopted a model for the accretion of metals onto dust-grains similar to \citet{popping2017}, based on the work by \citet{zhukovska2014}. One of the fundamental differences between the two models is that \citet{popping2017} calculate the timescale for metal accretion (among others) as a function of gas-phase metallicity (see also \citealt{li2019}), whereas \citet{vijayan2019} calculate this as a function of the dust-abundance of the dense ISM.  This can make a significant difference to the dust growth rate at early times. As a consequence, it takes longer for galaxies to reach the saturation limit in the \dtm, explaining the observed redshift evolution compared to for instance \citet{popping2017} and \citet{li2019}.


A clear difference in the implementation of dust physics in the various models is the use of yield tables to describe the dust enrichment of the ISM by AGB stars and supernovae. The \citet{popping2017}, \citet{li2019} and \citet{triani2020} models assume a fixed dust yield for all AGB stars and SNe, motivated by either observations or theoretical models. The \citet{vijayan2019} model assumes a fixed condensation fraction of metals into dust for SNe and uses yield tables for AGB stars. The \citet{hou2019} model adopts yield tables for dust enrichment by SNe, but ignores the effect of AGB stars. The \citet{graziani2019} model is the only model discussed in this work that adopts yield tables for dust enrichment by AGB stars and SNe.  The \citet{graziani2019} model best reproduces the $z>4$ observational constrains between \dtg\ and \dtm\ and metallicity, although this can not be fully ascribed to the use of yield tables as other processes such as the destruction of dust and the accretion of metals onto dust grains also play a relevant role. Nevertheless, including yield tables will be a relevant improvement for models that track the buildup of dust in galaxies, especially in the regimes where dust growth in the ISM is not yet a dominant contributor to the dust budget of galaxies such as low-metallicity environments and the early Universe.

\section{Discussion}
\label{sec:discussion}

\subsection{No cosmic evolution in galaxy dust properties with metallicity}
We found that the \dtg\ and \dtm\ are a strong function of galaxy properties as traced for instance by metallicity and that there is little evolution (at z<5.3) in this function, i.e., galaxies evolve along the relation between \dtg\ or \dtm\ and metallicity (Section \ref{Observation} and Figure \ref{fig:obs}). The latter implies that the dust-chemistry time-scales at play must be short enough for said relations to be in place within galaxies already at $z\sim5$, when the Universe was only $\sim 1.1$Gyr old. If grain growth through the accretion of metals onto dust grains is to play an important role in the early Universe for metal-enriched galaxies (e.g., \citet{Mancini2015}, \citet{popping2017}, c.f., \citealt{ferrara2016}) the grain-growth timescale must be significantly shorter than 1.1 Gyr. Furthermore, the lack of evolution also suggests that the balance between dust formation and destruction (as well as dust-poor inflows and dust-enriched outflows) is constant as a function of metallicity and therefore closely coupled to metal enrichment, i.e., the star-formation history rather than other external parameters \citep{Wendt2021}. 

Importantly, the relation between \dtg\ and metallicity has implications for studies of the cosmic evolution of cold gas based on dust mass measurements. These works infer the gas-mass of galaxies from their estimated dust mass observed in the far-infrared and mm domains. In recent years, single millimeter dust-continuum measurements in particular have been used as estimators of galaxy cold gas masses (see for example \citealt{scoville2017}, \citealt{kaasinen2019}). In some cases these recipes account for changes in the \dtg\ as a function of metallicity \citep{Bertemes2018,wang2022}, but not always. The robustness of such dust-based gas mass estimates will be improved by using the \dtg-metallicity relation tabulated in Table~\ref{tab:obs_fit}. These refined results will affect in particular the studies focusing on low-mass galaxies and galaxies during the epoch of reionisation.


\subsection{\dtg\ and \dtm\ as a constraint for models}
The tension between the predictions by the various models and the observational constraints from absorption studies demonstrate that the \dtg\ and \dtm\ of galaxies should be used as a stringent constraint for models that include dust chemistry. Indeed, none of the models succeeds in reproducing the derived values over the entire range in redshift and gas-phase metallicity now probed by the observations. Besides reproducing the exact shape of the various relations, some of the models also fail to reproduce the wide scatter observed. 

Galaxy formation models are complex indeed, including different recipes for all the baryonic physics that is relevant to describe the evolution of galaxies within their cosmological environments (e.g., the accretion and cooling of matter onto galaxies, the formation of stars, black hole growth, stellar and black hole feedback, metal enrichment, see the reviews by \citealt{somerville2015} and \citealt{naab2017}). As a consequence, it is sometimes complicated to understand why one model reproduces a certain observation better than another. Due to the close link between dust and gas metals in galaxies (dust originates from metals) the situation is more informative for the dust budget in galaxies compared to their metal content. This assumes that dust and metals follow each other closely in flows of gas, for example in galactic outflows.  We found that differences in the relation between \dtg\ or \dtm\ and gas-phase metallicity can be well explained by the different implementations of the dust-chemistry model in combination with the timescale for star formation (see Section \ref{sec:differences}). 

We find that the critical metallicity at which the \dtg\ and \dtm\ of galaxies rapidly increases predicted by the \citet{popping2017}, \citet{triani2020} and \citet{li2019} is too large at $z>2$. This is suggestive of a star-formation timescale in these simulations that is too short. Alternative approaches to contribute to bringing the models in better agreement with the data include unphysically high condensation efficiencies in stellar ejecta of 100 per cent (as explored in \citealt{popping2017}) or timescales for the accretion of metals onto dust grains that are unrealistically short. A heavily reduced efficiency of dust destruction by SN blast waves at lower metallicities could also contribute, but is in strong disagreement with models (\citealt{hu2019}, \citealt{Martinez-gonzalez2019},  see also the discussion in \citealt{Ferrara2021}), this may not contribute enough to resolve the tension (e.g., when turning dust-destruction off in the \citealt{popping2017} models the upturn in the \dtm\ occurs at metallicities only $\sim$0.2 dex lower than for the fiducial model with dust destruction, indicating that not enough dust is produced to begin with in low-metallicity galaxies or that the galaxies are enriched too quickly). What this ultimately demonstrates, is that the relations of galaxies' \dtg\ and \dtm\ with gas-phase metallicity provide a new and unique approach to not only constrain the dust-physics in galaxies, but also the timescale of star-formation at different cosmic epochs. 

\subsubsection{Caveats on the implementation of dust physics}
The physics of dust formation and destruction is in reality significantly more complex than can be captured in the various sub-grid recipes adopted in models and is also affected by various physical phenomena not included in the discussed cosmological simulations at all. In Section \ref{sec:differences} we attempted to highlight how the differences between model predictions can be explained by their implementations for the formation and destruction of dust. It is nevertheless relevant to take into account some of the current caveats or shortcomings in the implementations of dust-physics as these may also play a relevant role in shaping the evolution of dust in galaxies and the agreement with observations. Below we discuss a number of examples relevant for the presented works.

As a first example we can take the accretion of metals onto dust grains which is thought to take place in the cold ISM of galaxies. The models discussed in this work do not include the chemical networks to cool down the gas to typical temperatures and densities of the molecular ISM (well below the temperatures reached in cosmological simulations of 10000 K). Instead the models rely on sub-grid approaches to calculate the efficiency of metal accretion on dust grains and make sub-grid choices to decide in which fraction of the ISM of galaxies grain growth may take place (e.g., in a fixed dense gas fraction of 0.1 as implemented in \citealt{hou2019} or in star-forming SPH particles as implemented in \citet{graziani2019}). It is not yet clear if these choices are close to reality and the success of one over the other may therefore not necessarily imply a better physical description of the dust physics at play. A coarse description of the cooling of the ISM not only affects the efficiency of grain-growth but also the timescales for the formation of stars. The inclusion of chemical networks to better describe the temperature and density conditions of the ISM in simulations (e.g., \citealt{Kannan2021}) may therefore have a non-negligible impact on the formation of dust and the \dtg\ and \dtm\ of galaxies.

As a second example we can take the destruction of dust by supernovae. Dust destruction occurs close around supernovae when the reversed shock passes (e.g., \citealt{bocchio2016}) and at the scale of the ISM when the supernova shock wave passes through the ISM. The latter is implemented in the simulations (roughly following the approach from \citealt{mckee1989}, but the gas mass cleared of dust has a dependence on the density of the ISM \citep{temim2015} which is not included in the models. The former process is accounted for in the simulations by adopting effective yields for supernovae (corresponding to the dust condensation efficiency after the reversed shock passed). The true effective yield can in reality not be captured by a single number and will depend on a number of parameters not resolved in cosmological simulations. The systematic uncertainties associated to both destruction processes are likely also affecting the buildup of dust in galaxies.

As a last example, somewhat related to the first example, we can take the close coupling between radiation and dust. The exposure of grains to different types of radiation fields from different stars and at different wavelengths has a strong impact on grain surface chemistry (e.g., \citealt{ivlev2015}). The presence of charged dust-grains has an impact on the evolution of the multiphase ISM and affects the grain sizes \citep{glatzle2021}. These effects ultimately play a role in for example the efficiency of dust formation through the accretion of metals. More specifically, some works have discussed that metals accreted on the surface of dust grains in the coldest ISM (10-20 $K$) are captured in icy mantles. These mantles are thought to be photo-desorbed once they return to the diffuse ISM making grain-growth in the coldest ISM an inefficient channel for the formation of dust \citep{ferrara2016, ceccarelli2018}.

\subsection{The next steps}
This work includes the determination of the \dtg\ and \dtm\ of galaxies as obtained through absorption studies. By nature, these studies probe the \dtg\ and \dtm\ of a galaxy along an individual sightline in the neutral ISM, rather than measuring the integrated quantities. A complementary approach to study the \dtg\ and \dtm\ of galaxies would be to measure their dust and gas content and metallicity through emission studies. \citet{shapley2020} utilised ALMA dust continuum and CO observations to measure the dust content and gas content of 4 galaxies with metallicity measurements based on rest-optical emission lines, and found that the \dtg\ of galaxies at $z\sim2$ with solar metallicity is similar to the \dtg\ found in local galaxies with solar metallicity. Extending such observations to probe a broader dynamic range in gas-phase metallicity would be important to obtain an alternative measurement of the relation between metallicity and \dtg\ and probe the importance of systematics between various methodologies to measure the \dtg\ and \dtm\ of galaxies. Additionally, future modeling work could focus on mocking the absorption studies, drawing the metallicity, \dtg\ and \dtm\ of galaxies from individual sightlines in the modeled volume.

We have shown that measurements of the dust properties of galaxies through absorption studies provide important constraints for galaxy evolution models. Most models discussed in this work predict a change in slope in the relation between \dtg\ or \dtm\ with gas-phase metallicity which is not evident in the absorber data. A larger sample of observational constraints at $12 + \log{(O/H)} < 8$ is required to shed further light on this. Furthermore, a statistically more robust sample of absorbers at $z>4$ will provide key information about the redshift evolution of the discussed relations and prove to be essential to estimate the dust timescales necessary for galaxies to reach a balance in the dust-chemistry which leads to the non-evolution of those relations.

We also note that further studies looking at the global dust properties as a function of other galaxy physical properties would also be welcome. In particular, a detailed knowledge of the star formation rate and stellar mass associated with the quasar absorbers will enable a fairer comparison with some of the models presented here. Specifically, some of the model s-shape relations occur because the nature of the ISM of large main-sequence galaxies significantly
differs from dwarf galaxies. Indeed, these galaxy types show varying structure of ISM environments and are affected differently by chemical, radiative and mechanical feedback effects. In the future, a more detailed comparison will come from drawing galaxies through sightlines and even mocking their emission using radiative transfer codes (although the various assumptions that need to be made when applying radiative transfer codes do introduce additional systematic uncertainties).

In this work we focused on the dust-abundance of galaxies. Nevertheless, the depletion pattern of individual elements can be obtained through absorption studies \citep{mattsson2019} and captures key information on the type of dust present and their formation channels. For example, \citet{Dwek2016} argues that iron in galaxies is more depleted than can be explained by dust condensation in the ejecta of AGB stars and core-collapse supernova alone, suggesting that iron in the ISM depletes onto dust grains. A careful accounting of metal depletion in simulations compared to observations has the potential to test the composition of dust predicted by models and the contribution to the total dust budget by the various formation channels. This will push galaxy formation models to improve their dust chemistry recipes for individual elements, for instance by including dust yield tables for AGB stars and supernovae (e.g., \citealt{ferrarotti2006}, \citealt{Bianchi2007} and \citealt{mattsson2010}, see for instance \citealt{graziani2019}) rather than assuming fixed condensation efficiencies. 

Lastly, the disagreement between models and observations concering the \dtg\ and \dtm\ of galaxies as a function of their metallicity (see Figures~\ref{fig:dtg} and \ref{fig:dtm}) asks for a careful re-evaluation of the implementation in the discussed models of the production and destruction of dust in concert with the timescale for star-formation. This is especially relevant if similar type of comparisons based on drawing sightlines in the simulated volumes will result in a similar disagreement. The data and comparisons presented in this work provide a unique approach to breaking the degeneracy between dust-chemistry implementation and the timescale for star-formation, with significant consequences for the modeling of the buildup of stellar mass over cosmic time.

\section{Summary and conclusion}
\label{sec:summary}
In this work we presented a comprehensive overview of the relation between the metallicity and the dust-to-gas (\dtg) and dust-to-metal (\dtm) ratio of galaxies  from $z=0$ to $z=5$. In particular, we gathered the latest observational measurements from studies of luminous galaxies at $z=0$ \citep{de-vis2019}, from absorption studies covering a redshift range $0<z<5$ \citep{decia2016,wiseman2017,peroux2020}, and from a study probing the \dtg\ of luminous galaxies at $z\sim 2$ \citep{shapley2020}. The observations are compared to predictions by a number of semi-analytic \citep{popping2017, vijayan2019,triani2020} and hydrodynamical \citep{li2019,hou2019,graziani2019} models. Our main results include:
\begin{itemize}
    \item A linear relation exists between the \dtg\ and metallicity of the observed galaxies across cosmic time. Only at metallicities less than $12 + \log{O/H} < 8$ are there hints of an increased scatter. There is little to no evolution in the observed relation from $0<z<5$. Similarly, the relation between \dtm\ and gas-phase metallicity is also fairly constant with time. The lack of evolution in the relations between \dtg\ and \dtm\ and metallicity are indicative of a balance between the formation and destruction of dust, already at $z=5$ when the Universe was 1.2 Gyr old. The $z=0$ data shows a different slope, but we note that the dynamic range in metallicity covered at $z=0$ is narrower than at higher redshifts.
    \item None of the presented theoretical models is able to successfully reproduce the \dtg\ and \dtm\ of galaxies as a function of their metallicity over the entire redshift range considered in this work. 
     \item The observations suggest that the timescale for star-formation in the \citet{popping2017} and \citet{li2019} is too short, whereas the timescale for star-formation in \citet{vijayan2019} at $z<2$ is too long. The observations furthermore 
     suggest that too many metals are accreted onto dust grains at $z<2$ in the \citet{hou2019} model.
    \item Notable successes of the models include the good match between  \dtg\ and \dtm\ predicted by the \citet{graziani2019} model at $z>4$ and observations, the agreement between observations of the \dtg\ and \dtm\ for galaxies at $z>2$ with the predictions by \citet{vijayan2019} and \citet{hou2019} models and that the \citet{popping2017} and \citet{li2019} models reproduce at all redshifts the \dtg\ of galaxies with metallicities larger than $12 + \log{(O/H)}= 8$.
    \item Using the derived relation between \dtg\ and metallicy, the cold gas mass of galaxies can be more reliably estimated from their millimeter dust-continuum brightness or dust mass, taking metallicity dependencies into account.
\end{itemize}

The comparison of the observational and theoretical compilations of the \dtg\ and \dtm\ properties of galaxies demonstrates the power of dust-physics in constraining galaxy formation models. These results will be ever more constraining once the grain properties of dust (size distribution and composition) are better characterised through future optical and space- and ground-based NIR studies (e.g., from the James Webb Space Telescope and Extremely Large Telescope). A better understanding of the buildup of dust and their properties will have tremendous implications for galaxy formation and evolution as a whole, ranging from better constraining the timescales of star-formation as well as the buildup of dust in the ISM to shield UV radiation, to the effect that various types of dust grains have on the absorption and re-emission of stellar radiation.

\section*{Acknowledgements}
The authors thank Luca Graziani, Kuan-Chou Hou, Qi Li, Dian Triani and Aswin Vijayan for making available to us model predictions, providing insights into the workings of the models and providing comments on an earlier version of this paper. The authors also thank Ryan McKinnon, Mark Vogelsberger and Xuejian Shen for useful discussions regarding the \citet{mckinnon2018} model. This work initiated from a discussion with Alice Shapley and benefited from discussions with Felix Priestley, Desika Narayanan and Michele Ginolfi. The authors furthermore thank the anonymous referee for a constructive report.

\section*{Data availability}
The compilations of observations and model predictions that support the findings of this study are available on request from the corresponding author.


\bibliographystyle{mnras}
\bibliography{main.bib}

\begin{thebibliography}{}
\makeatletter
\relax
\def\mn@urlcharsother{\let\do\@makeother \do\$\do\&\do\#\do\^\do\_\do\%\do\~}
\def\mn@doi{\begingroup\mn@urlcharsother \@ifnextchar [ {\mn@doi@}
  {\mn@doi@[]}}
\def\mn@doi@[#1]#2{\def\@tempa{#1}\ifx\@tempa\@empty \href
  {http://dx.doi.org/#2} {doi:#2}\else \href {http://dx.doi.org/#2} {#1}\fi
  \endgroup}
\def\mn@eprint#1#2{\mn@eprint@#1:#2::\@nil}
\def\mn@eprint@arXiv#1{\href {http://arxiv.org/abs/#1} {{\tt arXiv:#1}}}
\def\mn@eprint@dblp#1{\href {http://dblp.uni-trier.de/rec/bibtex/#1.xml}
  {dblp:#1}}
\def\mn@eprint@#1:#2:#3:#4\@nil{\def\@tempa {#1}\def\@tempb {#2}\def\@tempc
  {#3}\ifx \@tempc \@empty \let \@tempc \@tempb \let \@tempb \@tempa \fi \ifx
  \@tempb \@empty \def\@tempb {arXiv}\fi \@ifundefined
  {mn@eprint@\@tempb}{\@tempb:\@tempc}{\expandafter \expandafter \csname
  mn@eprint@\@tempb\endcsname \expandafter{\@tempc}}}

\bibitem[\protect\citeauthoryear{{Aoyama}, {Hou}, {Shimizu}, {Hirashita},
  {Todoroki}, {Choi}  \& {Nagamine}}{{Aoyama} et~al.}{2017}]{aoyama2017}
{Aoyama} S.,  {Hou} K.-C.,  {Shimizu} I.,  {Hirashita} H.,  {Todoroki} K.,
  {Choi} J.-H.,   {Nagamine} K.,  2017, \mn@doi [\mnras]
  {10.1093/mnras/stw3061}, \href
  {https://ui.adsabs.harvard.edu/\#abs/2017MNRAS.466..105A} {466, 105}

\bibitem[\protect\citeauthoryear{{Asano}, {Takeuchi}, {Hirashita}  \&
  {Inoue}}{{Asano} et~al.}{2013}]{asano2013}
{Asano} R.~S.,  {Takeuchi} T.~T.,  {Hirashita} H.,   {Inoue} A.~K.,  2013,
  \mn@doi [Earth, Planets and Space] {10.5047/eps.2012.04.014}, \href
  {https://ui.adsabs.harvard.edu/abs/2013EP&S...65..213A} {65, 213}

\bibitem[\protect\citeauthoryear{{Augustin} et~al.,}{{Augustin}
  et~al.}{2018}]{augustin2018}
{Augustin} R.,  et~al., 2018, \mn@doi [\mnras] {10.1093/mnras/sty1287}, \href
  {https://ui.adsabs.harvard.edu/abs/2018MNRAS.478.3120A} {478, 3120}

\bibitem[\protect\citeauthoryear{{Berg} et~al.,}{{Berg}
  et~al.}{2021}]{berg2021}
{Berg} T. A.~M.,  et~al., 2021, \mn@doi [\mnras] {10.1093/mnras/stab184}, \href
  {https://ui.adsabs.harvard.edu/abs/2021MNRAS.502.4009B} {502, 4009}

\bibitem[\protect\citeauthoryear{{Bertemes} et~al.,}{{Bertemes}
  et~al.}{2018}]{Bertemes2018}
{Bertemes} C.,  et~al., 2018, \mn@doi [\mnras] {10.1093/mnras/sty963}, \href
  {https://ui.adsabs.harvard.edu/abs/2018MNRAS.478.1442B} {478, 1442}

\bibitem[\protect\citeauthoryear{{Bianchi} \& {Schneider}}{{Bianchi} \&
  {Schneider}}{2007}]{Bianchi2007}
{Bianchi} S.,  {Schneider} R.,  2007, \mn@doi [\mnras]
  {10.1111/j.1365-2966.2007.11829.x}, \href
  {https://ui.adsabs.harvard.edu/abs/2007MNRAS.378..973B} {378, 973}

\bibitem[\protect\citeauthoryear{{Blitz}}{{Blitz}}{1993}]{Blitz1993}
{Blitz} L.,  1993, in {Levy} E.~H.,  {Lunine} J.~I.,  eds, Protostars and
  Planets III. p.~125

\bibitem[\protect\citeauthoryear{{Blitz} \& {Rosolowsky}}{{Blitz} \&
  {Rosolowsky}}{2006}]{blitz2006}
{Blitz} L.,  {Rosolowsky} E.,  2006, \mn@doi [\apj] {10.1086/505417}, \href
  {https://ui.adsabs.harvard.edu/\#abs/2006ApJ...650..933B} {650, 933}

\bibitem[\protect\citeauthoryear{{Bocchio}, {Marassi}, {Schneider}, {Bianchi},
  {Limongi}  \& {Chieffi}}{{Bocchio} et~al.}{2016}]{bocchio2016}
{Bocchio} M.,  {Marassi} S.,  {Schneider} R.,  {Bianchi} S.,  {Limongi} M.,
  {Chieffi} A.,  2016, \mn@doi [\aap] {10.1051/0004-6361/201527432}, \href
  {https://ui.adsabs.harvard.edu/abs/2016A&A...587A.157B} {587, A157}

\bibitem[\protect\citeauthoryear{{Bouwens} et~al.,}{{Bouwens}
  et~al.}{2020}]{Bouwens2020}
{Bouwens} R.,  et~al., 2020, \mn@doi [\apj] {10.3847/1538-4357/abb830}, \href
  {https://ui.adsabs.harvard.edu/abs/2020ApJ...902..112B} {902, 112}

\bibitem[\protect\citeauthoryear{{Buat}, {Boselli}, {Gavazzi}  \&
  {Bonfanti}}{{Buat} et~al.}{2002}]{Buat2002}
{Buat} V.,  {Boselli} A.,  {Gavazzi} G.,   {Bonfanti} C.,  2002, \mn@doi [\aap]
  {10.1051/0004-6361:20011832}, \href
  {https://ui.adsabs.harvard.edu/abs/2002A&A...383..801B} {383, 801}

\bibitem[\protect\citeauthoryear{{Calzetti}, {Armus}, {Bohlin}, {Kinney},
  {Koornneef}  \& {Storchi-Bergmann}}{{Calzetti} et~al.}{2000}]{Calzetti2000}
{Calzetti} D.,  {Armus} L.,  {Bohlin} R.~C.,  {Kinney} A.~L.,  {Koornneef} J.,
   {Storchi-Bergmann} T.,  2000, \mn@doi [\apj] {10.1086/308692}, \href
  {https://ui.adsabs.harvard.edu/abs/2000ApJ...533..682C} {533, 682}

\bibitem[\protect\citeauthoryear{{Cazaux} \& {Spaans}}{{Cazaux} \&
  {Spaans}}{2009}]{Cazaux2009}
{Cazaux} S.,  {Spaans} M.,  2009, \mn@doi [\aap] {10.1051/0004-6361:200811302},
  \href {https://ui.adsabs.harvard.edu/abs/2009A&A...496..365C} {496, 365}

\bibitem[\protect\citeauthoryear{{Ceccarelli}, {Viti}, {Balucani}  \&
  {Taquet}}{{Ceccarelli} et~al.}{2018}]{ceccarelli2018}
{Ceccarelli} C.,  {Viti} S.,  {Balucani} N.,   {Taquet} V.,  2018, \mn@doi
  [\mnras] {10.1093/mnras/sty313}, \href
  {https://ui.adsabs.harvard.edu/abs/2018MNRAS.476.1371C} {476, 1371}

\bibitem[\protect\citeauthoryear{{Chabrier}}{{Chabrier}}{2003}]{chabrier2003}
{Chabrier} G.,  2003, \mn@doi [Publications of the Astronomical Society of the
  Pacific] {10.1086/376392}, \href
  {https://ui.adsabs.harvard.edu/\#abs/2003PASP..115..763C} {115, 763}

\bibitem[\protect\citeauthoryear{{Chiang}, {Sandstrom}, {Chastenet}, {Johnson},
  {Leroy}  \& {Utomo}}{{Chiang} et~al.}{2018}]{chiang2018}
{Chiang} I.-D.,  {Sandstrom} K.~M.,  {Chastenet} J.,  {Johnson} L.~C.,  {Leroy}
  A.~K.,   {Utomo} D.,  2018, \mn@doi [\apj] {10.3847/1538-4357/aadc5f}, \href
  {https://ui.adsabs.harvard.edu/abs/2018ApJ...865..117C} {865, 117}

\bibitem[\protect\citeauthoryear{{Chon}, {Omukai}  \& {Schneider}}{{Chon}
  et~al.}{2021}]{chon2021}
{Chon} S.,  {Omukai} K.,   {Schneider} R.,  2021, \mn@doi [\mnras]
  {10.1093/mnras/stab2497}, \href
  {https://ui.adsabs.harvard.edu/abs/2021MNRAS.508.4175C} {508, 4175}

\bibitem[\protect\citeauthoryear{{Croton} et~al.,}{{Croton}
  et~al.}{2016}]{croton2016}
{Croton} D.~J.,  et~al., 2016, \mn@doi [\apjs] {10.3847/0067-0049/222/2/22},
  \href {https://ui.adsabs.harvard.edu/abs/2016ApJS..222...22C} {222, 22}

\bibitem[\protect\citeauthoryear{{Dav{\'e}}, {Angl{\'e}s-Alc{\'a}zar},
  {Narayanan}, {Li}, {Rafieferantsoa}  \& {Appleby}}{{Dav{\'e}}
  et~al.}{2019}]{dave2019}
{Dav{\'e}} R.,  {Angl{\'e}s-Alc{\'a}zar} D.,  {Narayanan} D.,  {Li} Q.,
  {Rafieferantsoa} M.~H.,   {Appleby} S.,  2019, \mn@doi [\mnras]
  {10.1093/mnras/stz937}, \href
  {https://ui.adsabs.harvard.edu/abs/2019MNRAS.486.2827D} {486, 2827}

\bibitem[\protect\citeauthoryear{{De Cia}}{{De Cia}}{2018}]{decia2018a}
{De Cia} A.,  2018, \mn@doi [\aap] {10.1051/0004-6361/201833034}, \href
  {https://ui.adsabs.harvard.edu/#abs/2018A&A...613L...2D} {613, L2}

\bibitem[\protect\citeauthoryear{{De Cia}, {Ledoux}, {Savaglio}, {Schady}  \&
  {Vreeswijk}}{{De Cia} et~al.}{2013}]{decia2013}
{De Cia} A.,  {Ledoux} C.,  {Savaglio} S.,  {Schady} P.,   {Vreeswijk} P.~M.,
  2013, \mn@doi [\aap] {10.1051/0004-6361/201321834}, \href
  {https://ui.adsabs.harvard.edu/abs/2013A&A...560A..88D} {560, A88}

\bibitem[\protect\citeauthoryear{{De Cia}, {Ledoux}, {Mattsson}, {Petitjean},
  {Srianand}, {Gavignaud}  \& {Jenkins}}{{De Cia} et~al.}{2016}]{decia2016}
{De Cia} A.,  {Ledoux} C.,  {Mattsson} L.,  {Petitjean} P.,  {Srianand} R.,
  {Gavignaud} I.,   {Jenkins} E.~B.,  2016, \mn@doi [\aap]
  {10.1051/0004-6361/201527895}, \href
  {https://ui.adsabs.harvard.edu/#abs/2016A&A...596A..97D} {596, A97}

\bibitem[\protect\citeauthoryear{{De Cia}, {Ledoux}, {Petitjean}  \&
  {Savaglio}}{{De Cia} et~al.}{2018}]{decia2018}
{De Cia} A.,  {Ledoux} C.,  {Petitjean} P.,   {Savaglio} S.,  2018, \mn@doi
  [\aap] {10.1051/0004-6361/201731970}, \href
  {http://adsabs.harvard.edu/abs/2018A%26A...611A..76D} {611, A76}

\bibitem[\protect\citeauthoryear{{De Vis} et~al.,}{{De Vis}
  et~al.}{2019}]{de-vis2019}
{De Vis} P.,  et~al., 2019, \mn@doi [\aap] {10.1051/0004-6361/201834444}, \href
  {https://ui.adsabs.harvard.edu/\#abs/2019A&A...623A...5D} {623, A5}

\bibitem[\protect\citeauthoryear{{Draine}}{{Draine}}{1978}]{Draine1978}
{Draine} B.~T.,  1978, \mn@doi [\apjs] {10.1086/190513}, \href
  {https://ui.adsabs.harvard.edu/abs/1978ApJS...36..595D} {36, 595}

\bibitem[\protect\citeauthoryear{{Draine}}{{Draine}}{2003}]{draine2003}
{Draine} B.~T.,  2003, \mn@doi [Annual Review of Astronomy and Astrophysics]
  {10.1146/annurev.astro.41.011802.094840}, \href
  {https://ui.adsabs.harvard.edu/\#abs/2003ARA&A..41..241D} {41, 241}

\bibitem[\protect\citeauthoryear{{Draine} \& {Li}}{{Draine} \&
  {Li}}{2007}]{draine2007}
{Draine} B.~T.,  {Li} A.,  2007, \mn@doi [\apj] {10.1086/511055}, \href
  {https://ui.adsabs.harvard.edu/abs/2007ApJ...657..810D} {657, 810}

\bibitem[\protect\citeauthoryear{{Dutta} et~al.,}{{Dutta}
  et~al.}{2021}]{Dutta21b}
{Dutta} R.,  et~al., 2021, \mn@doi [\mnras] {10.1093/mnras/stab2752}, \href
  {https://ui.adsabs.harvard.edu/abs/2021MNRAS.508.4573D} {508, 4573}

\bibitem[\protect\citeauthoryear{{Dwek}}{{Dwek}}{1998}]{dwek1998}
{Dwek} E.,  1998, \mn@doi [\apj] {10.1086/305829}, \href
  {https://ui.adsabs.harvard.edu/abs/1998ApJ...501..643D} {501, 643}

\bibitem[\protect\citeauthoryear{{Dwek}}{{Dwek}}{2016}]{Dwek2016}
{Dwek} E.,  2016, \mn@doi [\apj] {10.3847/0004-637X/825/2/136}, \href
  {https://ui.adsabs.harvard.edu/abs/2016ApJ...825..136D} {825, 136}

\bibitem[\protect\citeauthoryear{{Dwek} \& {Scalo}}{{Dwek} \&
  {Scalo}}{1980}]{dwek1980}
{Dwek} E.,  {Scalo} J.~M.,  1980, \mn@doi [\apj] {10.1086/158100}, \href
  {https://ui.adsabs.harvard.edu/abs/1980ApJ...239..193D} {239, 193}

\bibitem[\protect\citeauthoryear{{Feldmann}}{{Feldmann}}{2015}]{feldmann2015}
{Feldmann} R.,  2015, \mn@doi [\mnras] {10.1093/mnras/stv552}, \href
  {https://ui.adsabs.harvard.edu/abs/2015MNRAS.449.3274F} {449, 3274}

\bibitem[\protect\citeauthoryear{{Ferrara} \& {Peroux}}{{Ferrara} \&
  {Peroux}}{2021}]{Ferrara2021}
{Ferrara} A.,  {Peroux} C.,  2021, \mn@doi [\mnras] {10.1093/mnras/stab761},
  \href {https://ui.adsabs.harvard.edu/abs/2021MNRAS.503.4537F} {503, 4537}

\bibitem[\protect\citeauthoryear{{Ferrara}, {Viti}  \& {Ceccarelli}}{{Ferrara}
  et~al.}{2016}]{ferrara2016}
{Ferrara} A.,  {Viti} S.,   {Ceccarelli} C.,  2016, \mn@doi [\mnras]
  {10.1093/mnrasl/slw165}, \href
  {https://ui.adsabs.harvard.edu/abs/2016MNRAS.463L.112F} {463, L112}

\bibitem[\protect\citeauthoryear{{Ferrarotti} \& {Gail}}{{Ferrarotti} \&
  {Gail}}{2006}]{ferrarotti2006}
{Ferrarotti} A.~S.,  {Gail} H.~P.,  2006, \mn@doi [\aap]
  {10.1051/0004-6361:20041198}, \href
  {https://ui.adsabs.harvard.edu/abs/2006A&A...447..553F} {447, 553}

\bibitem[\protect\citeauthoryear{{Fu}, {Kauffmann}, {Li}  \& {Guo}}{{Fu}
  et~al.}{2012}]{fu2012}
{Fu} J.,  {Kauffmann} G.,  {Li} C.,   {Guo} Q.,  2012, \mn@doi [\mnras]
  {10.1111/j.1365-2966.2012.21356.x}, \href
  {https://ui.adsabs.harvard.edu/abs/2012MNRAS.424.2701F} {424, 2701}

\bibitem[\protect\citeauthoryear{{Fudamoto} et~al.,}{{Fudamoto}
  et~al.}{2021}]{Fudamoto2021}
{Fudamoto} Y.,  et~al., 2021, \mn@doi [\nat] {10.1038/s41586-021-03846-z},
  \href {https://ui.adsabs.harvard.edu/abs/2021Natur.597..489F} {597, 489}

\bibitem[\protect\citeauthoryear{{Fukui} \& {Kawamura}}{{Fukui} \&
  {Kawamura}}{2010}]{Fukui2010}
{Fukui} Y.,  {Kawamura} A.,  2010, \mn@doi [\araa]
  {10.1146/annurev-astro-081309-130854}, \href
  {https://ui.adsabs.harvard.edu/abs/2010ARA&A..48..547F} {48, 547}

\bibitem[\protect\citeauthoryear{{Galametz}, {Madden}, {Galliano}, {Hony},
  {Bendo}  \& {Sauvage}}{{Galametz} et~al.}{2011}]{Galametz2011}
{Galametz} M.,  {Madden} S.~C.,  {Galliano} F.,  {Hony} S.,  {Bendo} G.~J.,
  {Sauvage} M.,  2011, \mn@doi [\aap] {10.1051/0004-6361/201014904}, \href
  {https://ui.adsabs.harvard.edu/abs/2011A&A...532A..56G} {532, A56}

\bibitem[\protect\citeauthoryear{{Galliano}, {Dwek}  \& {Chanial}}{{Galliano}
  et~al.}{2008}]{galliano2008}
{Galliano} F.,  {Dwek} E.,   {Chanial} P.,  2008, \mn@doi [\apj]
  {10.1086/523621}, \href
  {https://ui.adsabs.harvard.edu/abs/2008ApJ...672..214G} {672, 214}

\bibitem[\protect\citeauthoryear{{Galliano} et~al.,}{{Galliano}
  et~al.}{2021}]{galliano2021}
{Galliano} F.,  et~al., 2021, \mn@doi [\aap] {10.1051/0004-6361/202039701},
  \href {https://ui.adsabs.harvard.edu/abs/2021A&A...649A..18G} {649, A18}

\bibitem[\protect\citeauthoryear{{Giannetti} et~al.,}{{Giannetti}
  et~al.}{2017}]{giannetti2017}
{Giannetti} A.,  et~al., 2017, \mn@doi [\aap] {10.1051/0004-6361/201731728},
  \href {https://ui.adsabs.harvard.edu/abs/2017A&A...606L..12G} {606, L12}

\bibitem[\protect\citeauthoryear{{Ginolfi}, {Graziani}, {Schneider}, {Marassi},
  {Valiante}, {Dell'Agli}, {Ventura}  \& {Hunt}}{{Ginolfi}
  et~al.}{2018}]{Ginolfi2018}
{Ginolfi} M.,  {Graziani} L.,  {Schneider} R.,  {Marassi} S.,  {Valiante} R.,
  {Dell'Agli} F.,  {Ventura} P.,   {Hunt} L.~K.,  2018, \mn@doi [\mnras]
  {10.1093/mnras/stx2572}, \href
  {https://ui.adsabs.harvard.edu/abs/2018MNRAS.473.4538G} {473, 4538}

\bibitem[\protect\citeauthoryear{{Glatzle}, {Graziani}  \& {Ciardi}}{{Glatzle}
  et~al.}{2022}]{glatzle2021}
{Glatzle} M.,  {Graziani} L.,   {Ciardi} B.,  2022, \mn@doi [\mnras]
  {10.1093/mnras/stab3459}, \href
  {https://ui.adsabs.harvard.edu/abs/2022MNRAS.510.1068G} {510, 1068}

\bibitem[\protect\citeauthoryear{{Gnedin}}{{Gnedin}}{2010}]{gnedin2010}
{Gnedin} N.~Y.,  2010, \mn@doi [\apj] {10.1088/2041-8205/721/2/L79}, \href
  {https://ui.adsabs.harvard.edu/\#abs/2010ApJ...721L..79G} {721, L79}

\bibitem[\protect\citeauthoryear{{Goldsmith}}{{Goldsmith}}{2001}]{Goldsmith2001}
{Goldsmith} P.~F.,  2001, \mn@doi [\apj] {10.1086/322255}, \href
  {https://ui.adsabs.harvard.edu/abs/2001ApJ...557..736G} {557, 736}

\bibitem[\protect\citeauthoryear{{Gong}, {Ostriker}  \& {Wolfire}}{{Gong}
  et~al.}{2017}]{Gong2017}
{Gong} M.,  {Ostriker} E.~C.,   {Wolfire} M.~G.,  2017, \mn@doi [\apj]
  {10.3847/1538-4357/aa7561}, \href
  {https://ui.adsabs.harvard.edu/abs/2017ApJ...843...38G} {843, 38}

\bibitem[\protect\citeauthoryear{{Gould} \& {Salpeter}}{{Gould} \&
  {Salpeter}}{1963}]{Gould1963}
{Gould} R.~J.,  {Salpeter} E.~E.,  1963, \mn@doi [\apj] {10.1086/147654}, \href
  {https://ui.adsabs.harvard.edu/abs/1963ApJ...138..393G} {138, 393}

\bibitem[\protect\citeauthoryear{{Graziani}, {Schneider}, {Ginolfi}, {Hunt},
  {Maio}, {Glatzle}  \& {Ciardi}}{{Graziani} et~al.}{2020}]{graziani2019}
{Graziani} L.,  {Schneider} R.,  {Ginolfi} M.,  {Hunt} L.~K.,  {Maio} U.,
  {Glatzle} M.,   {Ciardi} B.,  2020, \mn@doi [\mnras] {10.1093/mnras/staa796},
  \href {https://ui.adsabs.harvard.edu/abs/2020MNRAS.494.1071G} {494, 1071}

\bibitem[\protect\citeauthoryear{{Gruppioni} et~al.,}{{Gruppioni}
  et~al.}{2020}]{Gruppioni2020}
{Gruppioni} C.,  et~al., 2020, \mn@doi [\aap] {10.1051/0004-6361/202038487},
  \href {https://ui.adsabs.harvard.edu/abs/2020A&A...643A...8G} {643, A8}

\bibitem[\protect\citeauthoryear{{Hamanowicz}, {Peroux}, {Zwaan}  \&
  {Rahmani}}{{Hamanowicz} et~al.}{2020}]{hamanowicz2020}
{Hamanowicz} A.,  {Peroux} C.,  {Zwaan} M.~A.,   {Rahmani} H. e.~a.,  2020,
  MNRAS, Submitted

\bibitem[\protect\citeauthoryear{{Henriques}, {White}, {Thomas}, {Angulo},
  {Guo}, {Lemson}, {Springel}  \& {Overzier}}{{Henriques}
  et~al.}{2015}]{henriques2015}
{Henriques} B. M.~B.,  {White} S. D.~M.,  {Thomas} P.~A.,  {Angulo} R.,  {Guo}
  Q.,  {Lemson} G.,  {Springel} V.,   {Overzier} R.,  2015, \mn@doi [\mnras]
  {10.1093/mnras/stv705}, \href
  {https://ui.adsabs.harvard.edu/abs/2015MNRAS.451.2663H} {451, 2663}

\bibitem[\protect\citeauthoryear{{Hirashita} \& {Kuo}}{{Hirashita} \&
  {Kuo}}{2011}]{hirashita2011}
{Hirashita} H.,  {Kuo} T.-M.,  2011, \mn@doi [\mnras]
  {10.1111/j.1365-2966.2011.19131.x}, \href
  {https://ui.adsabs.harvard.edu/abs/2011MNRAS.416.1340H} {416, 1340}

\bibitem[\protect\citeauthoryear{{Hirashita}, {Tajiri}  \&
  {Kamaya}}{{Hirashita} et~al.}{2002}]{Hirashita2002}
{Hirashita} H.,  {Tajiri} Y.~Y.,   {Kamaya} H.,  2002, \mn@doi [\aap]
  {10.1051/0004-6361:20020605}, \href
  {https://ui.adsabs.harvard.edu/abs/2002A&A...388..439H} {388, 439}

\bibitem[\protect\citeauthoryear{{Hollenbach} \& {Salpeter}}{{Hollenbach} \&
  {Salpeter}}{1971}]{Hollenbach1971}
{Hollenbach} D.,  {Salpeter} E.~E.,  1971, \mn@doi [\apj] {10.1086/150754},
  \href {https://ui.adsabs.harvard.edu/abs/1971ApJ...163..155H} {163, 155}

\bibitem[\protect\citeauthoryear{{Hollenbach}, {Kaufman}, {Neufeld}, {Wolfire}
  \& {Goicoechea}}{{Hollenbach} et~al.}{2012}]{Hollenbach2012}
{Hollenbach} D.,  {Kaufman} M.~J.,  {Neufeld} D.,  {Wolfire} M.,   {Goicoechea}
  J.~R.,  2012, \mn@doi [\apj] {10.1088/0004-637X/754/2/105}, \href
  {https://ui.adsabs.harvard.edu/abs/2012ApJ...754..105H} {754, 105}

\bibitem[\protect\citeauthoryear{{Hopkins}}{{Hopkins}}{2015}]{hopkins2015}
{Hopkins} P.~F.,  2015, \mn@doi [\mnras] {10.1093/mnras/stv195}, \href
  {https://ui.adsabs.harvard.edu/abs/2015MNRAS.450...53H} {450, 53}

\bibitem[\protect\citeauthoryear{{Hou}, {Hirashita}, {Nagamine}, {Aoyama}  \&
  {Shimizu}}{{Hou} et~al.}{2017}]{hou2017}
{Hou} K.-C.,  {Hirashita} H.,  {Nagamine} K.,  {Aoyama} S.,   {Shimizu} I.,
  2017, \mn@doi [\mnras] {10.1093/mnras/stx877}, \href
  {https://ui.adsabs.harvard.edu/abs/2017MNRAS.469..870H} {469, 870}

\bibitem[\protect\citeauthoryear{{Hou}, {Aoyama}, {Hirashita}, {Nagamine}  \&
  {Shimizu}}{{Hou} et~al.}{2019}]{hou2019}
{Hou} K.-C.,  {Aoyama} S.,  {Hirashita} H.,  {Nagamine} K.,   {Shimizu} I.,
  2019, \mn@doi [\mnras] {10.1093/mnras/stz121}, \href
  {https://ui.adsabs.harvard.edu/abs/2019MNRAS.485.1727H} {485, 1727}

\bibitem[\protect\citeauthoryear{{Hu} et~al.,}{{Hu} et~al.}{2019}]{hu2019}
{Hu} W.,  et~al., 2019, \mn@doi [\mnras] {10.1093/mnras/stz2038}, \href
  {https://ui.adsabs.harvard.edu/abs/2019MNRAS.489.1619H} {489, 1619}

\bibitem[\protect\citeauthoryear{{Ishiki} \& {Okamoto}}{{Ishiki} \&
  {Okamoto}}{2017}]{Ishiki2017}
{Ishiki} S.,  {Okamoto} T.,  2017, \mn@doi [\mnras] {10.1093/mnrasl/slw253},
  \href {https://ui.adsabs.harvard.edu/abs/2017MNRAS.466L.123I} {466, L123}

\bibitem[\protect\citeauthoryear{{Issa}, {MacLaren}  \& {Wolfendale}}{{Issa}
  et~al.}{1990}]{Issa1990}
{Issa} M.~R.,  {MacLaren} I.,   {Wolfendale} A.~W.,  1990, \aap, \href
  {https://ui.adsabs.harvard.edu/abs/1990A&A...236..237I} {236, 237}

\bibitem[\protect\citeauthoryear{{Ivlev}, {Padovani}, {Galli}  \&
  {Caselli}}{{Ivlev} et~al.}{2015}]{ivlev2015}
{Ivlev} A.~V.,  {Padovani} M.,  {Galli} D.,   {Caselli} P.,  2015, \mn@doi
  [\apj] {10.1088/0004-637X/812/2/135}, \href
  {https://ui.adsabs.harvard.edu/abs/2015ApJ...812..135I} {812, 135}

\bibitem[\protect\citeauthoryear{{Jenkins}}{{Jenkins}}{2009}]{jenkins2009}
{Jenkins} E.~B.,  2009, \mn@doi [\apj] {10.1088/0004-637X/700/2/1299}, \href
  {http://adsabs.harvard.edu/abs/2009ApJ...700.1299J} {700, 1299}

\bibitem[\protect\citeauthoryear{{Jenkins} \& {Wallerstein}}{{Jenkins} \&
  {Wallerstein}}{2017}]{jenkins2017}
{Jenkins} E.~B.,  {Wallerstein} G.,  2017, \mn@doi [\apj]
  {10.3847/1538-4357/aa64d4}, \href
  {https://ui.adsabs.harvard.edu/#abs/2017ApJ...838...85J} {838, 85}

\bibitem[\protect\citeauthoryear{{Kaasinen} et~al.,}{{Kaasinen}
  et~al.}{2019}]{kaasinen2019}
{Kaasinen} M.,  et~al., 2019, \mn@doi [\apj] {10.3847/1538-4357/ab253b}, \href
  {https://ui.adsabs.harvard.edu/abs/2019ApJ...880...15K} {880, 15}

\bibitem[\protect\citeauthoryear{{Kannan}, {Garaldi}, {Smith}, {Pakmor},
  {Springel}, {Vogelsberger}  \& {Hernquist}}{{Kannan}
  et~al.}{2021}]{Kannan2021}
{Kannan} R.,  {Garaldi} E.,  {Smith} A.,  {Pakmor} R.,  {Springel} V.,
  {Vogelsberger} M.,   {Hernquist} L.,  2021, arXiv e-prints, \href
  {https://ui.adsabs.harvard.edu/abs/2021arXiv211000584K} {p. arXiv:2110.00584}

\bibitem[\protect\citeauthoryear{{Kataoka}, {Okuzumi}, {Tanaka}  \&
  {Nomura}}{{Kataoka} et~al.}{2014}]{Kataoka2014}
{Kataoka} A.,  {Okuzumi} S.,  {Tanaka} H.,   {Nomura} H.,  2014, \mn@doi [\aap]
  {10.1051/0004-6361/201323199}, \href
  {https://ui.adsabs.harvard.edu/abs/2014A&A...568A..42K} {568, A42}

\bibitem[\protect\citeauthoryear{{Kennicutt} \& {Evans}}{{Kennicutt} \&
  {Evans}}{2012}]{Kennicutt2012}
{Kennicutt} R.~C.,  {Evans} N.~J.,  2012, \mn@doi [\araa]
  {10.1146/annurev-astro-081811-125610}, \href
  {https://ui.adsabs.harvard.edu/abs/2012ARA&A..50..531K} {50, 531}

\bibitem[\protect\citeauthoryear{{Kewley} \& {Ellison}}{{Kewley} \&
  {Ellison}}{2008}]{kewley2008}
{Kewley} L.~J.,  {Ellison} S.~L.,  2008, \mn@doi [\apj] {10.1086/587500}, \href
  {https://ui.adsabs.harvard.edu/\#abs/2008ApJ...681.1183K} {681, 1183}

\bibitem[\protect\citeauthoryear{{Kewley}, {Nicholls}  \&
  {Sutherland}}{{Kewley} et~al.}{2019}]{kewley2019}
{Kewley} L.~J.,  {Nicholls} D.~C.,   {Sutherland} R.~S.,  2019, \mn@doi [\araa]
  {10.1146/annurev-astro-081817-051832}, \href
  {https://ui.adsabs.harvard.edu/abs/2019ARA&A..57..511K} {57, 511}

\bibitem[\protect\citeauthoryear{{Krumholz}, {Leroy}  \& {McKee}}{{Krumholz}
  et~al.}{2011}]{Krumholz2011}
{Krumholz} M.~R.,  {Leroy} A.~K.,   {McKee} C.~F.,  2011, \mn@doi [\apj]
  {10.1088/0004-637X/731/1/25}, \href
  {https://ui.adsabs.harvard.edu/abs/2011ApJ...731...25K} {731, 25}

\bibitem[\protect\citeauthoryear{{Lagos}, {Baugh}, {Lacey}, {Benson}, {Kim}  \&
  {Power}}{{Lagos} et~al.}{2011}]{lagos2011}
{Lagos} C. D.~P.,  {Baugh} C.~M.,  {Lacey} C.~G.,  {Benson} A.~J.,  {Kim}
  H.-S.,   {Power} C.,  2011, \mn@doi [\mnras]
  {10.1111/j.1365-2966.2011.19583.x}, \href
  {https://ui.adsabs.harvard.edu/\#abs/2011MNRAS.418.1649L} {418, 1649}

\bibitem[\protect\citeauthoryear{{Larson}}{{Larson}}{2005}]{Larson2005}
{Larson} R.~B.,  2005, \mn@doi [\mnras] {10.1111/j.1365-2966.2005.08881.x},
  \href {https://ui.adsabs.harvard.edu/abs/2005MNRAS.359..211L} {359, 211}

\bibitem[\protect\citeauthoryear{{Li}, {Narayanan}  \& {Dav{\'e}}}{{Li}
  et~al.}{2019}]{li2019}
{Li} Q.,  {Narayanan} D.,   {Dav{\'e}} R.,  2019, \mn@doi [\mnras]
  {10.1093/mnras/stz2684}, \href
  {https://ui.adsabs.harvard.edu/abs/2019MNRAS.490.1425L} {490, 1425}

\bibitem[\protect\citeauthoryear{{Li}, {Narayanan}, {Torrey}, {Dav{\'e}}  \&
  {Vogelsberger}}{{Li} et~al.}{2020}]{li2020}
{Li} Q.,  {Narayanan} D.,  {Torrey} P.,  {Dav{\'e}} R.,   {Vogelsberger} M.,
  2020, arXiv e-prints, \href
  {https://ui.adsabs.harvard.edu/abs/2020arXiv201203978L} {p. arXiv:2012.03978}

\bibitem[\protect\citeauthoryear{{Liang} et~al.,}{{Liang}
  et~al.}{2019}]{Liang2019}
{Liang} L.,  et~al., 2019, \mn@doi [\mnras] {10.1093/mnras/stz2134}, \href
  {https://ui.adsabs.harvard.edu/abs/2019MNRAS.489.1397L} {489, 1397}

\bibitem[\protect\citeauthoryear{{Lianou}, {Barmby}, {Mosenkov}, {Lehnert}  \&
  {Karczewski}}{{Lianou} et~al.}{2019}]{lianou2019}
{Lianou} S.,  {Barmby} P.,  {Mosenkov} A.,  {Lehnert} M.,   {Karczewski} O.,
  2019, arXiv e-prints, \href
  {https://ui.adsabs.harvard.edu/abs/2019arXiv190602712L} {p. arXiv:1906.02712}

\bibitem[\protect\citeauthoryear{{Lisenfeld} \& {Ferrara}}{{Lisenfeld} \&
  {Ferrara}}{1998}]{lisenfeld1998}
{Lisenfeld} U.,  {Ferrara} A.,  1998, \mn@doi [\apj] {10.1086/305354}, \href
  {https://ui.adsabs.harvard.edu/abs/1998ApJ...496..145L} {496, 145}

\bibitem[\protect\citeauthoryear{{Madau} \& {Dickinson}}{{Madau} \&
  {Dickinson}}{2014}]{Madau2014}
{Madau} P.,  {Dickinson} M.,  2014, \mn@doi [\araa]
  {10.1146/annurev-astro-081811-125615}, \href
  {https://ui.adsabs.harvard.edu/abs/2014ARA&A..52..415M} {52, 415}

\bibitem[\protect\citeauthoryear{{Maio}, {Ciardi}, {Yoshida}, {Dolag}  \&
  {Tornatore}}{{Maio} et~al.}{2009}]{Maio2009}
{Maio} U.,  {Ciardi} B.,  {Yoshida} N.,  {Dolag} K.,   {Tornatore} L.,  2009,
  \mn@doi [\aap] {10.1051/0004-6361/200912234}, \href
  {https://ui.adsabs.harvard.edu/abs/2009A&A...503...25M} {503, 25}

\bibitem[\protect\citeauthoryear{{Maiolino} \& {Mannucci}}{{Maiolino} \&
  {Mannucci}}{2019}]{maiolino2019}
{Maiolino} R.,  {Mannucci} F.,  2019, \mn@doi [\aapr]
  {10.1007/s00159-018-0112-2}, \href
  {https://ui.adsabs.harvard.edu/abs/2019A&ARv..27....3M} {27, 3}

\bibitem[\protect\citeauthoryear{{Mancini}, {Schneider}, {Graziani},
  {Valiante}, {Dayal}, {Maio}, {Ciardi}  \& {Hunt}}{{Mancini}
  et~al.}{2015}]{Mancini2015}
{Mancini} M.,  {Schneider} R.,  {Graziani} L.,  {Valiante} R.,  {Dayal} P.,
  {Maio} U.,  {Ciardi} B.,   {Hunt} L.~K.,  2015, \mn@doi [\mnras]
  {10.1093/mnrasl/slv070}, \href
  {https://ui.adsabs.harvard.edu/abs/2015MNRAS.451L..70M} {451, L70}

\bibitem[\protect\citeauthoryear{{Mart{\'\i}nez-Gonz{\'a}lez}, {W{\"u}nsch},
  {Silich}, {Tenorio-Tagle}, {Palou{\v{s}}}  \&
  {Ferrara}}{{Mart{\'\i}nez-Gonz{\'a}lez} et~al.}{2019}]{Martinez-gonzalez2019}
{Mart{\'\i}nez-Gonz{\'a}lez} S.,  {W{\"u}nsch} R.,  {Silich} S.,
  {Tenorio-Tagle} G.,  {Palou{\v{s}}} J.,   {Ferrara} A.,  2019, \mn@doi [\apj]
  {10.3847/1538-4357/ab571b}, \href
  {https://ui.adsabs.harvard.edu/abs/2019ApJ...887..198M} {887, 198}

\bibitem[\protect\citeauthoryear{{Mattsson}, {Wahlin}  \&
  {H{\"o}fner}}{{Mattsson} et~al.}{2010}]{mattsson2010}
{Mattsson} L.,  {Wahlin} R.,   {H{\"o}fner} S.,  2010, \mn@doi [\aap]
  {10.1051/0004-6361/200912084}, \href
  {https://ui.adsabs.harvard.edu/abs/2010A&A...509A..14M} {509, A14}

\bibitem[\protect\citeauthoryear{{Mattsson}, {De Cia}, {Andersen}  \&
  {Petitjean}}{{Mattsson} et~al.}{2019}]{mattsson2019}
{Mattsson} L.,  {De Cia} A.,  {Andersen} A.~C.,   {Petitjean} P.,  2019,
  \mn@doi [\aap] {10.1051/0004-6361/201731482}, \href
  {https://ui.adsabs.harvard.edu/abs/2019A&A...624A.103M} {624, A103}

\bibitem[\protect\citeauthoryear{{McKee}}{{McKee}}{1989}]{mckee1989}
{McKee} C.,  1989, in {Allamandola} L.~J.,  {Tielens} A.~G.~G.~M.,  eds,  Vol.
  135, Interstellar Dust. p.~431

\bibitem[\protect\citeauthoryear{{McKee} \& {Krumholz}}{{McKee} \&
  {Krumholz}}{2010}]{mckee2009}
{McKee} C.~F.,  {Krumholz} M.~R.,  2010, \mn@doi [\apj]
  {10.1088/0004-637X/709/1/308}, \href
  {https://ui.adsabs.harvard.edu/abs/2010ApJ...709..308M} {709, 308}

\bibitem[\protect\citeauthoryear{{McKinnon}, {Vogelsberger}, {Torrey},
  {Marinacci}  \& {Kannan}}{{McKinnon} et~al.}{2018}]{mckinnon2018}
{McKinnon} R.,  {Vogelsberger} M.,  {Torrey} P.,  {Marinacci} F.,   {Kannan}
  R.,  2018, \mn@doi [\mnras] {10.1093/mnras/sty1248}, \href
  {https://ui.adsabs.harvard.edu/#abs/2018MNRAS.478.2851M} {478, 2851}

\bibitem[\protect\citeauthoryear{{M{\'e}nard}, {Scranton}, {Fukugita}  \&
  {Richards}}{{M{\'e}nard} et~al.}{2010}]{menard2010}
{M{\'e}nard} B.,  {Scranton} R.,  {Fukugita} M.,   {Richards} G.,  2010,
  \mn@doi [\mnras] {10.1111/j.1365-2966.2010.16486.x}, \href
  {https://ui.adsabs.harvard.edu/#abs/2010MNRAS.405.1025M} {405, 1025}

\bibitem[\protect\citeauthoryear{{Naab} \& {Ostriker}}{{Naab} \&
  {Ostriker}}{2017}]{naab2017}
{Naab} T.,  {Ostriker} J.~P.,  2017, \mn@doi [\araa]
  {10.1146/annurev-astro-081913-040019}, \href
  {https://ui.adsabs.harvard.edu/abs/2017ARA&A..55...59N} {55, 59}

\bibitem[\protect\citeauthoryear{{Nozawa}, {Kozasa}  \& {Habe}}{{Nozawa}
  et~al.}{2006}]{nozawa2006}
{Nozawa} T.,  {Kozasa} T.,   {Habe} A.,  2006, \mn@doi [\apj] {10.1086/505639},
  \href {https://ui.adsabs.harvard.edu/abs/2006ApJ...648..435N} {648, 435}

\bibitem[\protect\citeauthoryear{{Okuzumi}, {Tanaka}  \& {Sakagami}}{{Okuzumi}
  et~al.}{2009}]{Okuzumi2009}
{Okuzumi} S.,  {Tanaka} H.,   {Sakagami} M.-a.,  2009, \mn@doi [\apj]
  {10.1088/0004-637X/707/2/1247}, \href
  {https://ui.adsabs.harvard.edu/abs/2009ApJ...707.1247O} {707, 1247}

\bibitem[\protect\citeauthoryear{{Omukai}}{{Omukai}}{2000}]{Omukai2000}
{Omukai} K.,  2000, \mn@doi [\apj] {10.1086/308776}, \href
  {https://ui.adsabs.harvard.edu/abs/2000ApJ...534..809O} {534, 809}

\bibitem[\protect\citeauthoryear{{Omukai}, {Tsuribe}, {Schneider}  \&
  {Ferrara}}{{Omukai} et~al.}{2005}]{Omukai2005}
{Omukai} K.,  {Tsuribe} T.,  {Schneider} R.,   {Ferrara} A.,  2005, \mn@doi
  [\apj] {10.1086/429955}, \href
  {https://ui.adsabs.harvard.edu/abs/2005ApJ...626..627O} {626, 627}

\bibitem[\protect\citeauthoryear{{Ostriker} \& {Silk}}{{Ostriker} \&
  {Silk}}{1973}]{Ostriker1973}
{Ostriker} J.,  {Silk} J.,  1973, \mn@doi [\apjl] {10.1086/181301}, \href
  {https://ui.adsabs.harvard.edu/abs/1973ApJ...184L.113O} {184, L113}

\bibitem[\protect\citeauthoryear{{Peek}, {M{\'e}nard}  \& {Corrales}}{{Peek}
  et~al.}{2015}]{peek2015}
{Peek} J.~E.~G.,  {M{\'e}nard} B.,   {Corrales} L.,  2015, \mn@doi [\apj]
  {10.1088/0004-637X/813/1/7}, \href
  {https://ui.adsabs.harvard.edu/\#abs/2015ApJ...813....7P} {813, 7}

\bibitem[\protect\citeauthoryear{{Peeples}, {Werk}, {Tumlinson}, {Oppenheimer},
  {Prochaska}, {Katz}  \& {Weinberg}}{{Peeples} et~al.}{2014}]{peeples2014}
{Peeples} M.~S.,  {Werk} J.~K.,  {Tumlinson} J.,  {Oppenheimer} B.~D.,
  {Prochaska} J.~X.,  {Katz} N.,   {Weinberg} D.~H.,  2014, \mn@doi [\apj]
  {10.1088/0004-637X/786/1/54}, \href
  {http://adsabs.harvard.edu/abs/2014ApJ...786...54P} {786, 54}

\bibitem[\protect\citeauthoryear{{P{\'e}roux} \& {Howk}}{{P{\'e}roux} \&
  {Howk}}{2020}]{peroux2020}
{P{\'e}roux} C.,  {Howk} J.~C.,  2020, \mn@doi [\araa]
  {10.1146/annurev-astro-021820-120014}, \href
  {https://ui.adsabs.harvard.edu/abs/2020ARA&A..5821820P} {58, 363}

\bibitem[\protect\citeauthoryear{{P{\'e}roux}, {Petitjean}, {Aracil}  \&
  {Srianand}}{{P{\'e}roux} et~al.}{2002}]{peroux2002}
{P{\'e}roux} C.,  {Petitjean} P.,  {Aracil} B.,   {Srianand} R.,  2002, \mn@doi
  [\na] {10.1016/S1384-1076(02)00179-3}, \href
  {https://ui.adsabs.harvard.edu/abs/2002NewA....7..577P} {7, 577}

\bibitem[\protect\citeauthoryear{{P{\'e}roux}, {Dessauges-Zavadsky},
  {D'Odorico}, {Kim}  \& {McMahon}}{{P{\'e}roux} et~al.}{2007}]{peroux2007}
{P{\'e}roux} C.,  {Dessauges-Zavadsky} M.,  {D'Odorico} S.,  {Kim} T.-S.,
  {McMahon} R.~G.,  2007, \mn@doi [\mnras] {10.1111/j.1365-2966.2007.12235.x},
  \href {https://ui.adsabs.harvard.edu/\#abs/2007MNRAS.382..177P} {382, 177}

\bibitem[\protect\citeauthoryear{{P{\'e}roux} et~al.,}{{P{\'e}roux}
  et~al.}{2019}]{peroux2019}
{P{\'e}roux} C.,  et~al., 2019, \mn@doi [\mnras] {10.1093/mnras/stz202}, \href
  {https://ui.adsabs.harvard.edu/\#abs/2019MNRAS.485.1595P} {485, 1595}

\bibitem[\protect\citeauthoryear{{P{\'e}roux}, {Nelson}, {van de Voort},
  {Pillepich}, {Marinacci}, {Vogelsberger}  \& {Hernquist}}{{P{\'e}roux}
  et~al.}{2020}]{PerouxNelson2020}
{P{\'e}roux} C.,  {Nelson} D.,  {van de Voort} F.,  {Pillepich} A.,
  {Marinacci} F.,  {Vogelsberger} M.,   {Hernquist} L.,  2020, \mn@doi [\mnras]
  {10.1093/mnras/staa2888}, \href
  {https://ui.adsabs.harvard.edu/abs/2020MNRAS.499.2462P} {499, 2462}

\bibitem[\protect\citeauthoryear{{Pilyugin} \& {Grebel}}{{Pilyugin} \&
  {Grebel}}{2016}]{Pilyugin2016}
{Pilyugin} L.~S.,  {Grebel} E.~K.,  2016, \mn@doi [\mnras]
  {10.1093/mnras/stw238}, \href
  {https://ui.adsabs.harvard.edu/abs/2016MNRAS.457.3678P} {457, 3678}

\bibitem[\protect\citeauthoryear{{Popping}, {Somerville}  \&
  {Trager}}{{Popping} et~al.}{2014}]{popping2014}
{Popping} G.,  {Somerville} R.~S.,   {Trager} S.~C.,  2014, \mn@doi [\mnras]
  {10.1093/mnras/stu991}, \href
  {https://ui.adsabs.harvard.edu/\#abs/2014MNRAS.442.2398P} {442, 2398}

\bibitem[\protect\citeauthoryear{{Popping}, {Behroozi}  \& {Peeples}}{{Popping}
  et~al.}{2015}]{popping2015}
{Popping} G.,  {Behroozi} P.~S.,   {Peeples} M.~S.,  2015, \mn@doi [\mnras]
  {10.1093/mnras/stv318}, \href
  {https://ui.adsabs.harvard.edu/abs/2015MNRAS.449..477P} {449, 477}

\bibitem[\protect\citeauthoryear{{Popping}, {Somerville}  \&
  {Galametz}}{{Popping} et~al.}{2017}]{popping2017}
{Popping} G.,  {Somerville} R.~S.,   {Galametz} M.,  2017, \mn@doi [\mnras]
  {10.1093/mnras/stx1545}, \href
  {https://ui.adsabs.harvard.edu/\#abs/2017MNRAS.471.3152P} {471, 3152}

\bibitem[\protect\citeauthoryear{{Poudel}, {Hein{\"a}m{\"a}ki}, {Tempel},
  {Einasto}, {Lietzen}  \& {Nurmi}}{{Poudel} et~al.}{2017}]{poudel2017}
{Poudel} A.,  {Hein{\"a}m{\"a}ki} P.,  {Tempel} E.,  {Einasto} M.,  {Lietzen}
  H.,   {Nurmi} P.,  2017, \mn@doi [\aap] {10.1051/0004-6361/201629639}, \href
  {https://ui.adsabs.harvard.edu/\#abs/2017A&A...597A..86P} {597, A86}

\bibitem[\protect\citeauthoryear{{Quiret} et~al.,}{{Quiret}
  et~al.}{2016}]{quiret2016}
{Quiret} S.,  et~al., 2016, \mn@doi [\mnras] {10.1093/mnras/stw524}, \href
  {http://adsabs.harvard.edu/abs/2016MNRAS.458.4074Q} {458, 4074}

\bibitem[\protect\citeauthoryear{{Rahmani} et~al.,}{{Rahmani}
  et~al.}{2016}]{rahmani2016}
{Rahmani} H.,  et~al., 2016, \mn@doi [\mnras] {10.1093/mnras/stw1965}, \href
  {http://adsabs.harvard.edu/abs/2016MNRAS.463..980R} {463, 980}

\bibitem[\protect\citeauthoryear{{R{\'e}my-Ruyer} et~al.,}{{R{\'e}my-Ruyer}
  et~al.}{2014}]{remy2014}
{R{\'e}my-Ruyer} A.,  et~al., 2014, \mn@doi [\aap]
  {10.1051/0004-6361/201322803}, \href
  {https://ui.adsabs.harvard.edu/abs/2014A&A...563A..31R} {563, A31}

\bibitem[\protect\citeauthoryear{{Roman-Duval} et~al.,}{{Roman-Duval}
  et~al.}{2014}]{roman-duval2014}
{Roman-Duval} J.,  et~al., 2014, \mn@doi [\apj] {10.1088/0004-637X/797/2/86},
  \href {https://ui.adsabs.harvard.edu/abs/2014ApJ...797...86R} {797, 86}

\bibitem[\protect\citeauthoryear{{Roman-Duval} et~al.,}{{Roman-Duval}
  et~al.}{2019}]{roman-duval2019}
{Roman-Duval} J.,  et~al., 2019, \mn@doi [\apj] {10.3847/1538-4357/aaf8bb},
  \href {https://ui.adsabs.harvard.edu/abs/2019ApJ...871..151R} {871, 151}

\bibitem[\protect\citeauthoryear{{Roman-Duval} et~al.,}{{Roman-Duval}
  et~al.}{2021}]{roman-duval2021}
{Roman-Duval} J.,  et~al., 2021, \mn@doi [\apj] {10.3847/1538-4357/abdeb6},
  \href {https://ui.adsabs.harvard.edu/abs/2021ApJ...910...95R} {910, 95}

\bibitem[\protect\citeauthoryear{{Roman-Duval} et~al.,}{{Roman-Duval}
  et~al.}{2022}]{roman-duval2022}
{Roman-Duval} J.,  et~al., 2022, arXiv e-prints, \href
  {https://ui.adsabs.harvard.edu/abs/2022arXiv220204765R} {p. arXiv:2202.04765}

\bibitem[\protect\citeauthoryear{{Romano}, {Nagamine}  \& {Hirashita}}{{Romano}
  et~al.}{2022}]{Romano2022}
{Romano} L. E.~C.,  {Nagamine} K.,   {Hirashita} H.,  2022, arXiv e-prints,
  \href {https://ui.adsabs.harvard.edu/abs/2022arXiv220205521R} {p.
  arXiv:2202.05521}

\bibitem[\protect\citeauthoryear{{Salim} \& {Narayanan}}{{Salim} \&
  {Narayanan}}{2020}]{Salim2020}
{Salim} S.,  {Narayanan} D.,  2020, \mn@doi [\araa]
  {10.1146/annurev-astro-032620-021933}, \href
  {https://ui.adsabs.harvard.edu/abs/2020ARA&A..58..529S} {58, 529}

\bibitem[\protect\citeauthoryear{{Schneider}, {Ferrara}  \&
  {Salvaterra}}{{Schneider} et~al.}{2004}]{Schneider2004}
{Schneider} R.,  {Ferrara} A.,   {Salvaterra} R.,  2004, \mn@doi [\mnras]
  {10.1111/j.1365-2966.2004.07876.x}, \href
  {https://ui.adsabs.harvard.edu/abs/2004MNRAS.351.1379S} {351, 1379}

\bibitem[\protect\citeauthoryear{{Schneider}, {Omukai}, {Inoue}  \&
  {Ferrara}}{{Schneider} et~al.}{2006}]{Schneider2006}
{Schneider} R.,  {Omukai} K.,  {Inoue} A.~K.,   {Ferrara} A.,  2006, \mn@doi
  [\mnras] {10.1111/j.1365-2966.2006.10391.x}, \href
  {https://ui.adsabs.harvard.edu/abs/2006MNRAS.369.1437S} {369, 1437}

\bibitem[\protect\citeauthoryear{{Scoville} et~al.,}{{Scoville}
  et~al.}{2017}]{scoville2017}
{Scoville} N.,  et~al., 2017, \mn@doi [\apj] {10.3847/1538-4357/aa61a0}, \href
  {https://ui.adsabs.harvard.edu/abs/2017ApJ...837..150S} {837, 150}

\bibitem[\protect\citeauthoryear{{Scudder}, {Ellison}, {El Meddah El Idrissi}
  \& {Poetrodjojo}}{{Scudder} et~al.}{2021}]{scudder2021}
{Scudder} J.~M.,  {Ellison} S.~L.,  {El Meddah El Idrissi} L.,   {Poetrodjojo}
  H.,  2021, \mn@doi [\mnras] {10.1093/mnras/stab2339}, \href
  {https://ui.adsabs.harvard.edu/abs/2021MNRAS.507.2468S} {507, 2468}

\bibitem[\protect\citeauthoryear{{Shapley}, {Cullen}, {Dunlop}, {McLure},
  {Kriek}, {Reddy}  \& {Sanders}}{{Shapley} et~al.}{2020}]{shapley2020}
{Shapley} A.~E.,  {Cullen} F.,  {Dunlop} J.~S.,  {McLure} R.~J.,  {Kriek} M.,
  {Reddy} N.~A.,   {Sanders} R.~L.,  2020, \mn@doi [\apjl]
  {10.3847/2041-8213/abc006}, \href
  {https://ui.adsabs.harvard.edu/abs/2020ApJ...903L..16S} {903, L16}

\bibitem[\protect\citeauthoryear{{Somerville} \& {Primack}}{{Somerville} \&
  {Primack}}{1999}]{somerville1999}
{Somerville} R.~S.,  {Primack} J.~R.,  1999, \mn@doi [\mnras]
  {10.1046/j.1365-8711.1999.03032.x}, \href
  {https://ui.adsabs.harvard.edu/abs/1999MNRAS.310.1087S} {310, 1087}

\bibitem[\protect\citeauthoryear{{Somerville}, {Hopkins}, {Cox}, {Robertson}
  \& {Hernquist}}{{Somerville} et~al.}{2008}]{somerville2008}
{Somerville} R.~S.,  {Hopkins} P.~F.,  {Cox} T.~J.,  {Robertson} B.~E.,
  {Hernquist} L.,  2008, \mn@doi [\mnras] {10.1111/j.1365-2966.2008.13805.x},
  \href {https://ui.adsabs.harvard.edu/abs/2008MNRAS.391..481S} {391, 481}

\bibitem[\protect\citeauthoryear{{Somerville}, {Popping}  \&
  {Trager}}{{Somerville} et~al.}{2015}]{somerville2015}
{Somerville} R.~S.,  {Popping} G.,   {Trager} S.~C.,  2015, \mn@doi [\mnras]
  {10.1093/mnras/stv1877}, \href
  {https://ui.adsabs.harvard.edu/abs/2015MNRAS.453.4337S} {453, 4337}

\bibitem[\protect\citeauthoryear{{Sparre} et~al.,}{{Sparre}
  et~al.}{2014}]{sparre2014}
{Sparre} M.,  et~al., 2014, \mn@doi [\apj] {10.1088/0004-637X/785/2/150}, \href
  {https://ui.adsabs.harvard.edu/abs/2014ApJ...785..150S} {785, 150}

\bibitem[\protect\citeauthoryear{{Springel}, {Di Matteo}  \&
  {Hernquist}}{{Springel} et~al.}{2005a}]{springel2005b}
{Springel} V.,  {Di Matteo} T.,   {Hernquist} L.,  2005a, \mn@doi [\mnras]
  {10.1111/j.1365-2966.2005.09238.x}, \href
  {https://ui.adsabs.harvard.edu/abs/2005MNRAS.361..776S} {361, 776}

\bibitem[\protect\citeauthoryear{{Springel} et~al.,}{{Springel}
  et~al.}{2005b}]{springel2005}
{Springel} V.,  et~al., 2005b, \mn@doi [\nat] {10.1038/nature03597}, \href
  {http://cdsads.u-strasbg.fr/abs/2005Natur.435..629S} {435, 629}

\bibitem[\protect\citeauthoryear{{Steidel}, {Adelberger}, {Giavalisco},
  {Dickinson}  \& {Pettini}}{{Steidel} et~al.}{1999}]{Steidel1999}
{Steidel} C.~C.,  {Adelberger} K.~L.,  {Giavalisco} M.,  {Dickinson} M.,
  {Pettini} M.,  1999, \mn@doi [\apj] {10.1086/307363}, \href
  {https://ui.adsabs.harvard.edu/abs/1999ApJ...519....1S} {519, 1}

\bibitem[\protect\citeauthoryear{{Takeuchi}, {Buat}, {Heinis}, {Giovannoli},
  {Yuan}, {Iglesias-P{\'a}ramo}, {Murata}  \& {Burgarella}}{{Takeuchi}
  et~al.}{2010}]{Takeuchi2010}
{Takeuchi} T.~T.,  {Buat} V.,  {Heinis} S.,  {Giovannoli} E.,  {Yuan} F.~T.,
  {Iglesias-P{\'a}ramo} J.,  {Murata} K.~L.,   {Burgarella} D.,  2010, \mn@doi
  [\aap] {10.1051/0004-6361/200913476}, \href
  {https://ui.adsabs.harvard.edu/abs/2010A&A...514A...4T} {514, A4}

\bibitem[\protect\citeauthoryear{{Temim}, {Dwek}, {Tchernyshyov}, {Boyer},
  {Meixner}, {Gall}  \& {Roman-Duval}}{{Temim} et~al.}{2015}]{temim2015}
{Temim} T.,  {Dwek} E.,  {Tchernyshyov} K.,  {Boyer} M.~L.,  {Meixner} M.,
  {Gall} C.,   {Roman-Duval} J.,  2015, \mn@doi [\apj]
  {10.1088/0004-637X/799/2/158}, \href
  {https://ui.adsabs.harvard.edu/abs/2015ApJ...799..158T} {799, 158}

\bibitem[\protect\citeauthoryear{{Tornatore}, {Ferrara}  \&
  {Schneider}}{{Tornatore} et~al.}{2007a}]{Tornatore2007b}
{Tornatore} L.,  {Ferrara} A.,   {Schneider} R.,  2007a, \mn@doi [\mnras]
  {10.1111/j.1365-2966.2007.12215.x}, \href
  {https://ui.adsabs.harvard.edu/abs/2007MNRAS.382..945T} {382, 945}

\bibitem[\protect\citeauthoryear{{Tornatore}, {Borgani}, {Dolag}  \&
  {Matteucci}}{{Tornatore} et~al.}{2007b}]{Tornatore2007a}
{Tornatore} L.,  {Borgani} S.,  {Dolag} K.,   {Matteucci} F.,  2007b, \mn@doi
  [\mnras] {10.1111/j.1365-2966.2007.12070.x}, \href
  {https://ui.adsabs.harvard.edu/abs/2007MNRAS.382.1050T} {382, 1050}

\bibitem[\protect\citeauthoryear{{Triani}, {Sinha}, {Croton}, {Pacifici}  \&
  {Dwek}}{{Triani} et~al.}{2020}]{triani2020}
{Triani} D.~P.,  {Sinha} M.,  {Croton} D.~J.,  {Pacifici} C.,   {Dwek} E.,
  2020, \mn@doi [\mnras] {10.1093/mnras/staa446}, \href
  {https://ui.adsabs.harvard.edu/abs/2020MNRAS.493.2490T} {493, 2490}

\bibitem[\protect\citeauthoryear{{Vijayan}, {Clay}, {Thomas}, {Yates},
  {Wilkins}  \& {Henriques}}{{Vijayan} et~al.}{2019}]{vijayan2019}
{Vijayan} A.~P.,  {Clay} S.~J.,  {Thomas} P.~A.,  {Yates} R.~M.,  {Wilkins}
  S.~M.,   {Henriques} B.~M.,  2019, \mn@doi [\mnras] {10.1093/mnras/stz1948},
  \href {https://ui.adsabs.harvard.edu/abs/2019MNRAS.489.4072V} {489, 4072}

\bibitem[\protect\citeauthoryear{{V{\'\i}lchez}, {Rela{\~n}o}, {Kennicutt}, {De
  Looze}, {Moll{\'a}}  \& {Galametz}}{{V{\'\i}lchez}
  et~al.}{2019}]{vilchez2019}
{V{\'\i}lchez} J.~M.,  {Rela{\~n}o} M.,  {Kennicutt} R.,  {De Looze} I.,
  {Moll{\'a}} M.,   {Galametz} M.,  2019, \mn@doi [\mnras]
  {10.1093/mnras/sty3455}, \href
  {https://ui.adsabs.harvard.edu/abs/2019MNRAS.483.4968V} {483, 4968}

\bibitem[\protect\citeauthoryear{{Vladilo}}{{Vladilo}}{2004}]{vladilo2004}
{Vladilo} G.,  2004, \mn@doi [\aap] {10.1051/0004-6361:20035897}, \href
  {https://ui.adsabs.harvard.edu/#abs/2004A&A...421..479V} {421, 479}

\bibitem[\protect\citeauthoryear{{Vogelsberger}, {McKinnon}, {O'Neil},
  {Marinacci}, {Torrey}  \& {Kannan}}{{Vogelsberger}
  et~al.}{2019}]{Vogelsberger2019}
{Vogelsberger} M.,  {McKinnon} R.,  {O'Neil} S.,  {Marinacci} F.,  {Torrey} P.,
    {Kannan} R.,  2019, \mn@doi [\mnras] {10.1093/mnras/stz1644}, \href
  {https://ui.adsabs.harvard.edu/abs/2019MNRAS.487.4870V} {487, 4870}

\bibitem[\protect\citeauthoryear{{Walter} et~al.,}{{Walter}
  et~al.}{2020}]{Walter2020}
{Walter} F.,  et~al., 2020, \mn@doi [\apj] {10.3847/1538-4357/abb82e}, \href
  {https://ui.adsabs.harvard.edu/abs/2020ApJ...902..111W} {902, 111}

\bibitem[\protect\citeauthoryear{{Wang} et~al.,}{{Wang}
  et~al.}{2022}]{wang2022}
{Wang} T.-M.,  et~al., 2022, arXiv e-prints, \href
  {https://ui.adsabs.harvard.edu/abs/2022arXiv220112070W} {p. arXiv:2201.12070}

\bibitem[\protect\citeauthoryear{{Wendt}, {Bouch{\'e}}, {Zabl}, {Schroetter}
  \& {Muzahid}}{{Wendt} et~al.}{2021}]{Wendt2021}
{Wendt} M.,  {Bouch{\'e}} N.~F.,  {Zabl} J.,  {Schroetter} I.,   {Muzahid} S.,
  2021, \mn@doi [\mnras] {10.1093/mnras/stab049}, \href
  {https://ui.adsabs.harvard.edu/abs/2021MNRAS.502.3733W} {502, 3733}

\bibitem[\protect\citeauthoryear{{Whitworth}, {Boffin}, {Watkins}  \&
  {Francis}}{{Whitworth} et~al.}{1998}]{Whitworth1998}
{Whitworth} A.,  {Boffin} H.,  {Watkins} S.,   {Francis} N.,  1998, Astronomy
  and Geophysics, \href {https://ui.adsabs.harvard.edu/abs/1998A&G....39f..10W}
  {39, 10}

\bibitem[\protect\citeauthoryear{{Wiseman}, {Schady}, {Bolmer}, {Kr{\"u}hler},
  {Yates}, {Greiner}  \& {Fynbo}}{{Wiseman} et~al.}{2017}]{wiseman2017}
{Wiseman} P.,  {Schady} P.,  {Bolmer} J.,  {Kr{\"u}hler} T.,  {Yates} R.~M.,
  {Greiner} J.,   {Fynbo} J.~P.~U.,  2017, \mn@doi [\aap]
  {10.1051/0004-6361/201629228}, \href
  {https://ui.adsabs.harvard.edu/#abs/2017A&A...599A..24W} {599, A24}

\bibitem[\protect\citeauthoryear{{Yamasawa}, {Habe}, {Kozasa}, {Nozawa},
  {Hirashita}, {Umeda}  \& {Nomoto}}{{Yamasawa} et~al.}{2011}]{yamasawa2011}
{Yamasawa} D.,  {Habe} A.,  {Kozasa} T.,  {Nozawa} T.,  {Hirashita} H.,
  {Umeda} H.,   {Nomoto} K.,  2011, \mn@doi [\apj]
  {10.1088/0004-637X/735/1/44}, \href
  {https://ui.adsabs.harvard.edu/abs/2011ApJ...735...44Y} {735, 44}

\bibitem[\protect\citeauthoryear{Zafar, Popping  \& P\'eroux}{Zafar
  et~al.}{2013}]{zafar2013a}
Zafar T.,  Popping A.,   P\'eroux C.,  2013, A\&A, 556, 140

\bibitem[\protect\citeauthoryear{{Zhukovska}}{{Zhukovska}}{2014}]{zhukovska2014}
{Zhukovska} S.,  2014, \mn@doi [\aap] {10.1051/0004-6361/201322989}, \href
  {https://ui.adsabs.harvard.edu/abs/2014A&A...562A..76Z} {562, A76}

\bibitem[\protect\citeauthoryear{{Zhukovska}, {Gail}  \&
  {Trieloff}}{{Zhukovska} et~al.}{2008}]{Zhukovska2008}
{Zhukovska} S.,  {Gail} H.~P.,   {Trieloff} M.,  2008, \mn@doi [\aap]
  {10.1051/0004-6361:20077789}, \href
  {https://ui.adsabs.harvard.edu/abs/2008A&A...479..453Z} {479, 453}

\bibitem[\protect\citeauthoryear{{de Bennassuti}, {Schneider}, {Valiante}  \&
  {Salvadori}}{{de Bennassuti} et~al.}{2014}]{Bennassuti2014}
{de Bennassuti} M.,  {Schneider} R.,  {Valiante} R.,   {Salvadori} S.,  2014,
  \mn@doi [\mnras] {10.1093/mnras/stu1962}, \href
  {https://ui.adsabs.harvard.edu/abs/2014MNRAS.445.3039D} {445, 3039}

\makeatother
\end{thebibliography}

\end{document}